\documentclass{mn2e}
\usepackage[dvips]{graphicx}
\usepackage{amssymb}

\begin{document}

\title[Properties of hydrodynamical galaxies]{Photometric and clustering properties of 
hydrodynamical galaxies in a cosmological volume: results at $z=0$}
\author[S. E. Nuza, K. Dolag \& A. Saro]{Sebasti\'an E. Nuza$^{1}$\thanks{E-mail: sebasn@mpa-garching.mpg.de}, Klaus Dolag$^{1}$ and Alexandro Saro$^{2,3}$\\
\\
$^1$ Max-Planck Institute for Astrophysics, Karl-Schwarzchild Str. 1, D85748, Garching, Germany\\
$^2$ Dipartamento di Astronomi\`a dell'Universita di Trieste, via Tiepolo 11, I-34131 Trieste, Italy\\
$^3$ INFN - National Institute for Nuclear Physics, Trieste, Italy \\
}

\maketitle

\begin{abstract}

  In this work, we present results for the photometric and clustering
  properties of galaxies that arise in a $\Lambda$ cold dark matter hydrodynamical simulation of 
  the local universe. The present-day distribution of matter was constructed to match 
  the observed large scale pattern of the {\it IRAS} 1.2-Jy galaxy survey.
  Our simulation follows the formation and evolution of galaxies in a cosmological 
  sphere with a volume of $\sim130^3$ $h^{-3}$ Mpc$^3$ including supernova feedback, 
  galactic winds, photoheating due to an uniform meta-galactic background and chemical enrichment 
  of the gas and stellar populations. However, we do not consider AGNs. 
  In the simulation, a total of $\sim 20000$ galaxies are formed above 
  the resolution limit, and around $60$ haloes are more massive than $\sim10^{14}$ M$_{\odot}$.
  Luminosities of the galaxies are calculated based on a stellar population synthesis model including 
  the attenuation by dust, which is calculated from the cold gas left within the simulated galaxies. 
  Environmental effects like colour bi-modality and differential clustering power of the hydrodynamical 
  galaxies are qualitatively similar to observed trends. Nevertheless, the overcooling present in the simulations 
  lead to too blue and overluminous brightest cluster galaxies (BCGs). To overcome this, we mimic the late-time suppression of star 
  formation in massive halos by ignoring recently formed stars with the aid of a simple post-processing recipe. 
  In this way we find luminosity functions, both for field and group/cluster galaxies, in better agreement with 
  observations. Specifically, the BCGs then follow the observed luminosity-halo mass relation. However, in such a case, 
  the colour bi-modality is basically lost, pointing towards a more complex interplay of late suppression of star formation 
  than what is given by the simple scheme adopted.

\end{abstract}


\section{Introduction}
\label{intro}

The observational study of galaxy populations has seen an 
outstanding progress in recent years. With the advent of 
large galaxy redshift surveys, such as the Sloan Digital Sky Survey 
(SDSS; York et al. 2000) and the Two-degree Field Galaxy Redshift 
Survey (2dFGRS; Colless et al. 2001) it was possible to extend our 
knowledge of the local Universe to a new level of accuracy. 

\begin{figure*}
\begin{center}
{\includegraphics[width=0.8\textwidth]{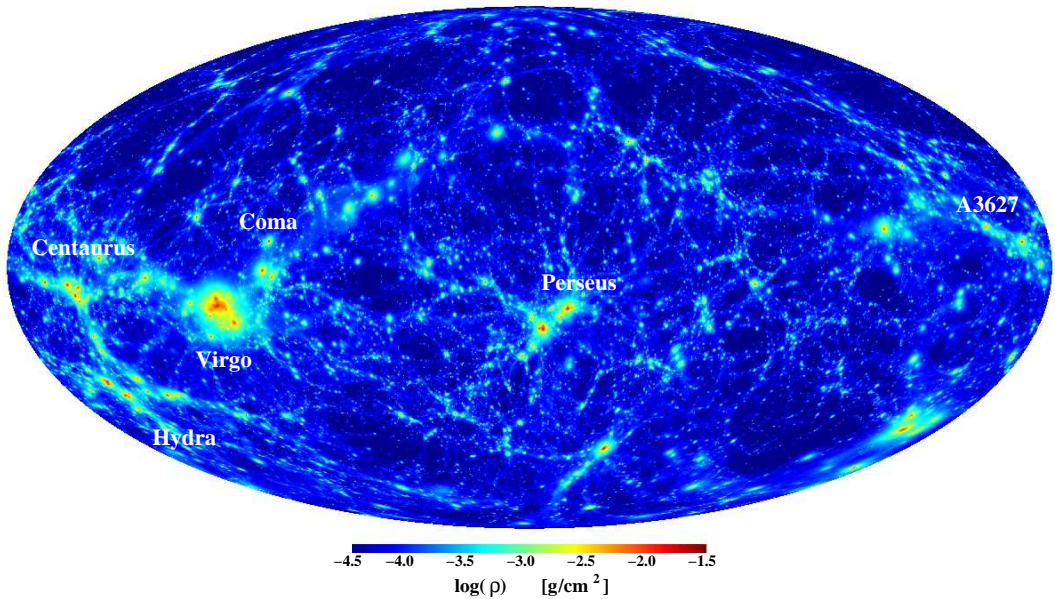}
\caption{Full-sky map of the simulated local volume in supergalactic coordinates. The map displays the 
projected gas density distribution up to a distance of $\sim 80$ $h^{-1}$ Mpc from the observer where main 
galaxy clusters are shown. Note that Virgo cluster, being the closest one to the observer, is particularly prominent.}
}
\end{center}
\label{full_sky}
\end{figure*}

In particular, it has been possible to carry out a robust determination of the 
luminosity function (LF) for galaxies in the field (Norberg et al. 2002; 
Blanton et al. 2003) in different spectral bands and to better establish it for 
those galaxies populating denser environments, such as groups and 
clusters (e.g. Popesso et al. 2004, 2006). Consistently with previous work 
(e.g. Lin et al. 1996; Colless et al. 1999) it has been found that the field LF is well described by a single 
Schechter function requiring, however, a higher value for the luminosity density 
of the Universe indicating that previous surveys suffered from selection effects and systematics due 
to photometry (Blanton et al. 2001). For higher density environments, and by means of the RASS-SDSS galaxy cluster 
survey, Popesso et al. (2004, 2006) found that a double Schechter function better fits the composite cluster 
LF accounting for a possible upturn in the number of faint objects as the luminosity of member galaxies 
decrease.

Moreover, the large number of galaxies with measured spectroscopic 
redshifts, compared to that obtained in the past, has made possible 
the determination of their two-point clustering properties in a reliable way out to scales of 
order of tens of Mpc. These studies have shown that the real, projected and redshift-space 
correlations can be well described by a decreasing power-law function that depends on the 
sample colour, having a correlation length that increases with absolute magnitude
(e.g. Norberg et al. 2001, 2002; Zehavi et al. 2002; Madgwick et al. 2003; Hawkins et al. 2003).

Both the LFs and the clustering properties of the galaxies are tools 
of fundamental importance in the study of galaxy formation since they provide a way to describe 
the most basic galaxy statistics. A successful model for structure formation must account 
for these observations.

From the theoretical point of view, the widely spread {\it semi-analytic} models (SAMs) of galaxy formation 
(e.g. Kauffmann et al. 1999; Springel et al. 2001; Mathis et al. 2002; 
De Lucia, Kauffmann \& White 2004; Springel et al. 2005; Bower et al. 2006; Cattaneo et al. 2006; 
Croton et al. 2006; Lagos, Cora \& Padilla 2008; Fontanot et al. 2009; Guo \& White 2009) provide a way to study the properties 
of galaxy populations with the advantage of a rapid exploration of the parameter space. In this approach, the galaxy population 
is followed within the skeleton provided by a parent dark matter simulation with the aid of physically 
motivated recipes to describe the different baryonic processes involved. These studies have pointed out the 
need of limiting excessive gas condensation in massive haloes to avoid the formation of very bright 
central galaxies (also known as brightest cluster galaxies; BCGs) that are inconsistent with observation. 
The usually invoked channel responsible 
for the star formation (SF) quenching in massive haloes is the active galactic nucleus (AGN) phenomenon. 
Similarly, it has been noted that in order to prevent an excess luminosity in the faint end of the galaxy 
field LF, a relatively strong SN feedback would be needed in smaller systems.

On the other hand, cosmological hydrodynamical simulations of galaxy samples, without including the 
effects of AGN feedback, have also been used to study the building up of the structure and their resulting 
properties in periodic boxes (e.g. Pearce et al. 2001; White et al. 2001; Yoshikawa et al. 2001; 
Oppenheimer \& Dav\'e 2006, 2008; Dav\'e \& Oppenheimer 2007; Ocvirk et al. 2008). 
In a similar way, using the dubbed {\it zooming} technique, several authors simulated hydrodynamical 
galaxies within high-resolution regions in a cosmological framework 
(e.g. Dolag et al. 2005; Saro et al. 2006, 2008, 2009; Crain et al. 2009). In particular, Saro et al. (2006) 
resimulated a set of galaxy clusters studying the composite cluster LF and the environmental 
induced galaxy properties in these high density regions. These authors show that it is possible to 
reproduce the general observed trends of the cluster galaxy population, including, e.g. the 
colour-magnitude relation, the age and colour cluster-centric distance dependence and the cluster LF 
(although the bright-end is affected by the presence of very bright BCGs). 

\begin{figure}
{\includegraphics[width=85mm]{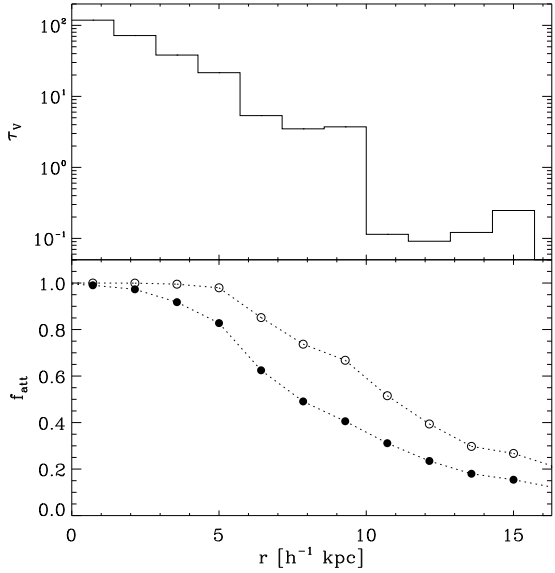}}
\caption{Estimated optical depth profile in the $V$-band for one of the most massive 
BCGs in the simulation including extended intra-cluster light ($M_*\sim10^{13}$ M$_{\sun}$) averaged 
over radial projected bins (top panel) and its correspondingly smoothed attenuation fraction due to dust (bottom panel) 
for the $b_{\rm j}$- and $K$-bands (open and filled circles respectively).}
\label{tauv}
\end{figure}

The present work makes use of the same {\it zooming} technique to simulate the observed local volume following 
the chemical enrichment of gas and stars until the present epoch. This 
enables us to estimate in a consistent 
way the luminosities and colours of galaxies formed in hydrodynamical simulations. 
The simulated volume is big enough to compute LFs, 
both for field and group/cluster galaxies, as well as galaxy correlation functions. 
In order to infer the way feedback acts in massive haloes (e.g. AGN feedback), we use a similar 
approach as SAMs, applying a simple recipe to the post-processed data of the simulation at $z=0$. 
In this way, we ignore a fraction of the late-formed stars which would not appear in such a scenario 
and study the resulting effects on the luminosity-dependent properties of the simulated galaxies.
It is important to note that, within the simple post-processing scheme adopted here, 
the simulations are no longer fully self-consistent, as also stars which are quenched by our post-processing
procedure still interact with the surrounding medium. However, this effect is not expected to change our
results significantly and can only be accessed with the next generation of hydrodynamical simulations including
sub-scale models for AGN feedback more directly.

The paper is organized as follows. In Section~\ref{simul} we describe the hydrodynamical 
cosmological simulation, together with the method used to compute galaxy luminosities and 
the associated dust-obscuration. 
In Section~\ref{results} we show the main results, presenting luminosities (for field 
and group/cluster galaxies), galaxy colours and correlation functions. We also discuss 
the implemented recipe to {\it suppress} SF in massive haloes.  
Finally, we close the paper with a summary and our conclusions in Section~\ref{concl}. 

\begin{table} 
\begin{center}
\caption{Simple SF suppression recipe: $z_{\rm cut}=(V_{\rm max}-V_0)/V_1$, where 
$z_{\rm cut}$ represents the redshift since which we do not consider star formation 
in a given galaxy. Systems having $V_{\rm max}<V_0$ are not affected by the suppression.}
\vspace{0.1cm}
\label{table1}
\begin{tabular}{cccc}
\hline
\hline
Model & $V_0$ & $V_1$\\
      & [km s$^{-1}$] &   [km s$^{-1}$]\\
\hline
1   & 100 & 150 \\
2   & 100 & 200 \\
3   & 100 & 250 \\
4   & 150 & 150 \\
5   & 150 & 200 \\
6   & 150 & 250 \\
7   & 200 & 150 \\
8   & 200 & 200 \\
9   & 200 & 250 \\
\hline
\hline
\end{tabular}
\end{center}
\end{table}

\section{The simulation}
\label{simul}

The cosmological simulation analyzed in this paper was generated to reproduce 
the large scale distribution of matter in the local universe within the context of a flat $\Lambda$CDM scenario. 
The cosmological parameters at present time are a matter density parameter $\Omega_{\rm m}=0.3$, a baryon density parameter 
$\Omega_{\rm b}=0.04$, a Hubble constant $H_0=100~h$ km s$^{-1}$ Mpc$^{-1}$ with $h=0.7$, and a rms density fluctuation 
$\sigma_8=0.9$, which correspond to the Wilkinson Microwave Anisotropy Probe (WMAP) 1 year best-fitting cosmology 
(Spergel et al. 2003).

In order to reproduce the local matter density field, the galaxy distribution in the {\it IRAS} 1.2-Jy galaxy survey (Fisher et al. 1994, 1995) was 
Gaussianly smoothed on a scale of 7 Mpc and then linearly evolved back in time up to $z=50$ with the method proposed by Kolatt et al. (1996). The 
resulting high redshift field is then used as a Gaussian constraint (Hoffman \& Ribak 1991) to assign the unperturbed positions of $2\times 51$ million 
gas and dark matter particles that are arranged on a glass-like distribution (e.g. Baugh, Gazta\~naga \& Efstathiou 1995). The mass of gas and dark 
matter particles is set to $4.7\times10^8$ and $3.1\times10^9$ $h^{-1}$ M$_{\odot}$ respectively. Every gas particle is able to 
produce three generations of stars, which results in a star particle mass of $\sim 1.6\times10^8$ $h^{-1}$ M$_{\odot}$.

We follow the evolution of the collisionless and gas components within 
a sphere with a comoving diameter of $160$ $h^{-1}$ Mpc that is embedded in a periodic cosmological box of $340$ $h^{-1}$ Mpc on a side. 
This makes the volume of the simulated universe a factor of $\sim 2$ times larger than in previous studies 
(e.g. Pearce et al. 2001). The region outside the sphere is sampled with $\sim 7$ million low resolution dark matter particles, that allow 
us to asses the effect of long-range gravitational tidal forces. The comoving softening length for gravitational forces is set 
to 7 $h^{-1}$ kpc (Plummer equivalent), value that is similar to the average particle separation in the cores of the densest simulated 
clusters. In Fig. \ref{full_sky}, we show the projected gas density distribution of the simulated local sphere. 

\begin{figure*}
{\includegraphics[width=85mm]{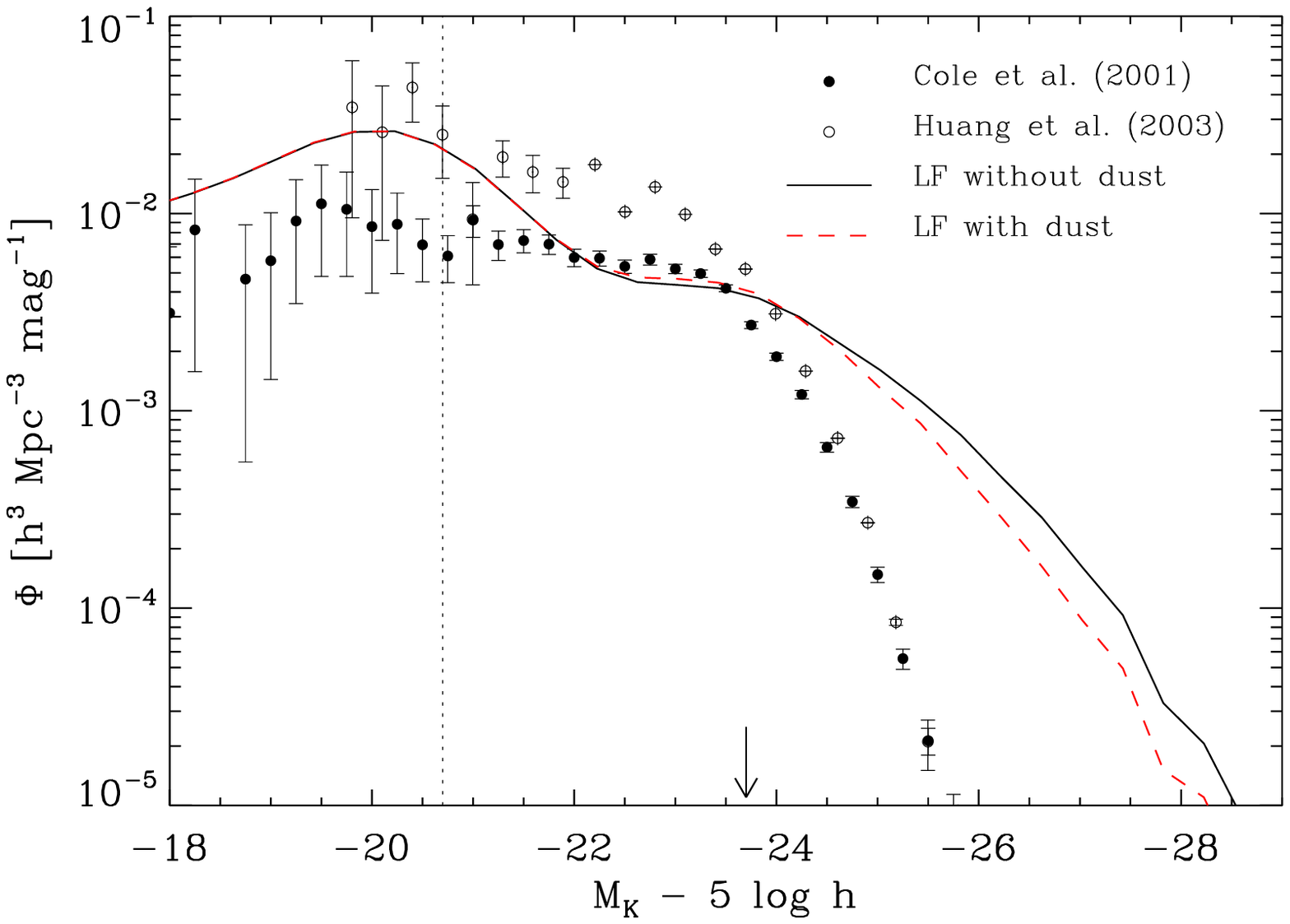}}
{\includegraphics[width=85mm]{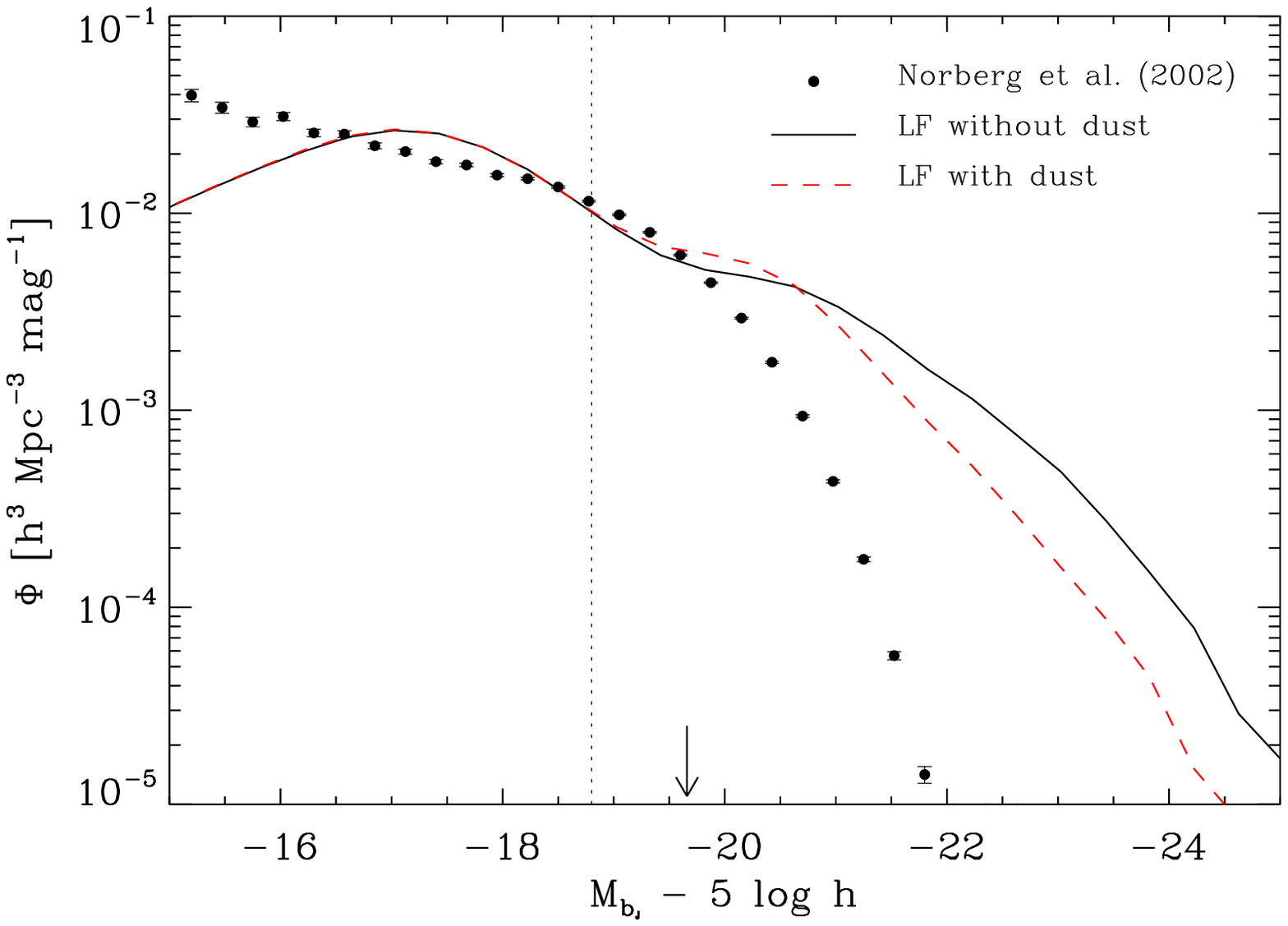}}
\caption{LFs for the simulated hydrodynamical galaxies at $z=0$ in comparison with the observations of Cole et al. (2001) 
and Huang et al. (2003) for the $K$-band (left-hand panel) and Norberg et al. (2002) for the $b_{\rm j}$-band (right-hand panel). The simulated 
LFs are renormalized to match observations at $M_* - 5 \log h$ (indicated by an arrow; see text). Dashed and solid lines show the resulting LFs for 
the dust-attenuated and non-attenuated cases respectively. The approximate resolution limit of the simulation in each band is indicated by the 
vertical dotted lines. 
}
\label{field}
\end{figure*}

The simulation was run using {\small GADGET-2} (Springel \& Hernquist 2002, 2003; Springel 2005) which is a 
Tree-smoothed particle hydrodynamics (SPH) code that fully conservates entropy during the evolution of the gas component, 
taking into account radiative cooling, heating by an ultraviolet (UV) meta-galactic background to emulate the reionization era 
(Haardt \& Madau 1996), and a subresolution scheme to treat SF, SN feedback and 
galactic winds. In particular, the set of parameters of the phenomenological wind model was chosen to obtain a escape 
velocity of $480$ km s$^{-1}$. For the SF recipe we assume that each star particle of the simulation is a 
single stellar population (SSP) where the relative number of stars with different masses is obtained by means of the initial mass function (IMF), which 
in this work is that of Salpeter (1955) normalized in the 0.1--100 M$_{\sun}$ mass range.

The metal content of gas the and stellar components is followed using the model presented in 
Tornatore et al. (2004; see also Tornatore et al. 2007). This scheme accounts for stellar evolution assessing the effect 
of the generation of Type II and Type Ia SN events (SNII and SNIa respectively), as well as of massive stars ending up their 
life in the asymptotic giant branch phase. These events release energy and metals to the surrounding medium according 
to the stellar lifetimes of the different populations (Maeder \& Meynet 1989) and to the SNIa rate 
(Matteucci \& Recchi 2001). It is assumed that energy feedback is only provided by SNs 
($10^{51}$ erg per event), while metals are distributed in all cases following the production 
of Fe, O, C, Si, Mg and S using the stellar yields found in Recchi, Matteucci \& D'Ercole (2001). 
In addition, the local cooling rate for each gas particle is self-consistently determined from its estimated 
[Fe/H] abundance and temperature using the tables given by Sutherland \& Dopita (1993). 
At each time step, the network of primordial species is solved assuming that 
photoionization only acts on them. In this regard, we expect that metal photoionization will have only a 
minor impact at the resolution covered by our simulation.

Bound systems at the present time are identified using the {\small SUBFIND} algorithm 
(e.g. Springel et al. 2001; Dolag et al. 2009). Firstly, haloes having more 
than 32 dark matter particles are selected using the friends-of-friends (FoF) algorithm with a linking 
length of 0.2 in units of the mean interparticle separation. Then, a total of 20, self bound, 
particles (ignoring gas particles) is set as a lower limit to identify substructures present 
in each FoF group that may be elegible as galaxies. As the detection of substructures 
is difficult at the low mass-end, we consider as numerically {\it resolved} galaxies only  
substructures in the simulation having more than 32 star particles, which results in a stellar mass 
resolution limit of $\sim 7\times10^{9}$ M$_{\odot}$.

\subsection{Luminosity of galaxies}
\label{lum_gal}

\subsubsection{Stellar population synthesis}
\label{SPS}

Similar to Saro et al. (2006), we compute the luminosities of our simulated galaxies 
using the stellar population synthesis model 
of Bruzual \& Charlot (2003) in different spectral bands. As mentioned in the previous section, 
each star particle is treated as a SSP with a formation time that corresponds to redshift $z_i$ 
and metallicity content $Z_i$. For every galaxy, this enable us to sum up the luminosity 
contributions $L_i$ of each stellar population in the following way

\begin{equation}
\label{Llambda}
L(\lambda)=\sum_i L_i(\lambda,z_i,Z_i){\rm e}^{-\tau_{\lambda}}
\end{equation}

\noindent where $\lambda$ is the wavelength of radiation, e$^{-\tau_{\lambda}}$ is the extinction factor due to the 
presence of dust (see e.g. Bruzual \& Charlot 2003), and $\tau_{\lambda}$ is the wavelength-dependent optical 
depth of the obscuring medium (see next section).

To compute the luminosities we use only stars inside the galactocentric distance that contains 83\% of the 
galaxy's baryonic mass, the so-called {\it optical radius} (e.g. Nuza et al. 2007). We have checked that 
adopting this condition has a negligible effect on the final absolute magnitudes.

\begin{figure}
{\includegraphics[width=80mm]{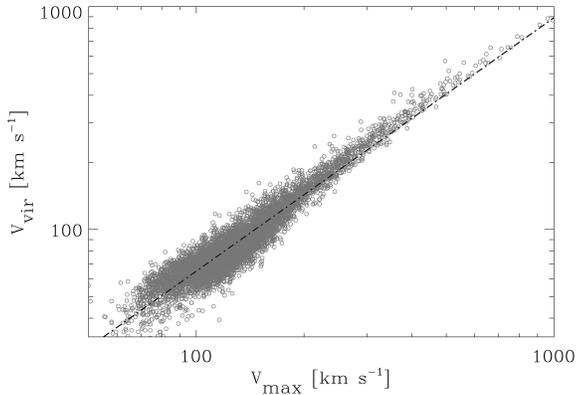}}
\caption{Halo virial velocity as a function of halo maximum circular velocity in the simulation. 
The dot-dashed line corresponds to the best linear fit.}
\label{vmax_vvir}
\end{figure}

\subsubsection{Attenuation by dust}
\label{dust}

In order to asses the effect of dust in the luminosity of galaxies we apply a similar procedure 
to Saro et al. (2009) to our hydrodynamical galaxies. The extinction model assumed is the one 
of Charlot \& Fall (2000). These authors considered a scenario where the radiation of newly born 
stars ($t\leq10^7$ yr) is attenuated by the presence of dust in their birth cloud. Older stellar populations 
are able to migrate out of the molecular clouds and, as a consequence, are mainly affected by the 
ambient ISM. Therefore, we assume the following age-dependent extinction curve

\begin{equation}
\label{taulambda}
\frac{\tau_{\lambda}}{\tau_V}= 
\left\{ \begin{array}{ll}
\left(\frac{\lambda}{550~{\rm nm}}\right)^{-0.7} & t\leq 10^{7} {\rm yr} \\
&\\
\mu\left(\frac{\lambda}{550~{\rm nm}}\right)^{-0.7} & t>10^7 {\rm yr}
\end{array} \right.
\end{equation}

\noindent where $\tau_V$ is the optical depth in the $V$-band, $\lambda$ is the wavelength of radiation, and $\mu$ is the fraction of the effective optical depth due to the ISM. In this paper we take $\mu$ randomly from a 
Gaussian distribution centered in 0.3 with a width of 0.2 and truncated at 0.1 and 1 in a similar way as done in previous works (e.g. De Lucia \& Blaizot 2007). 

Following Guiderdoni \& Rocca-Volmerage (1987) we assume that $\tau_V$ is proportional to the column density of cold gas in the line of sight having mean metallicity $Z_{\rm g}$. In particular, for the wavelengths of interest here, we can write

\begin{equation} 
\tau_V = \left(\frac{Z_{\rm g}}{Z_{\sun}}\right)^s \left(\frac{N_{\rm H}}{2.1 \times 10^{21}~{\rm cm^{-2}}}\right),
\end{equation} 

\noindent where $N_{\rm H}$ is the hydrogen column density and the mean gas metallicity exponent $s$ is equal 
to $1.6$ (valid for $\lambda > 200$ nm; see Guiderdoni \& Rocca-Volmerage for details). 

To estimate the spatial dependence of $\tau_V$ for our simulated galaxies we project their gas content onto a plane 
(say $xy$) with the aim at averaging the cold gas distribution as a function of projected 
distance to the centre. Using the obtained $\tau_V(r)$ radial profile we compute the differential dust-attenuated 
luminosity contributed by every star particle in a given point $(x,y)$ using its $z$-position as a proxy of its obscuration level, i.e. proportional to the amount of intervening cold gas between them and the galaxy outskirts. As a consequence, star particles behind the parent galaxy receive the total attenuation corresponding to its radial position $\tau_V(r)$, while those ahead are almost not attenuated at all. The luminosity of stellar populations with intermediate $z$-positions is accordingly affected.

As an example, in the top panel of Fig. \ref{tauv} we show the visible optical depth averaged over projected radial bins in 
one of the most massive galaxies present in the simulation. As stated above, this quantity depend both on metallicity and column density of cold gas present in the system, hence, the visible optical depth tends to be higher in the inner regions of our hydrodynamical galaxies. This is due to the fact that, in the simulation, the galactic central regions tend to have higher column densities with highly enriched gas. In the bottom panel, we plot the fraction of luminosity attenuation for star particles in both $b_{\rm j}$- and $K$-bands as a function of central distance (open and filled circles respectively). As expected, it can be seen that the attenuation correlates with the optical depth profile and that it is stronger for bluer bands.

Finally, we integrate over all stellar populations in order to estimate a global corrected magnitude for each 
galaxy in the volume (see Eq. \ref{Llambda}). In the rest of the paper we will always use {\it dust-attenuated} quantities unless explicitly stated. 

\begin{figure*}
{\includegraphics[width=85mm]{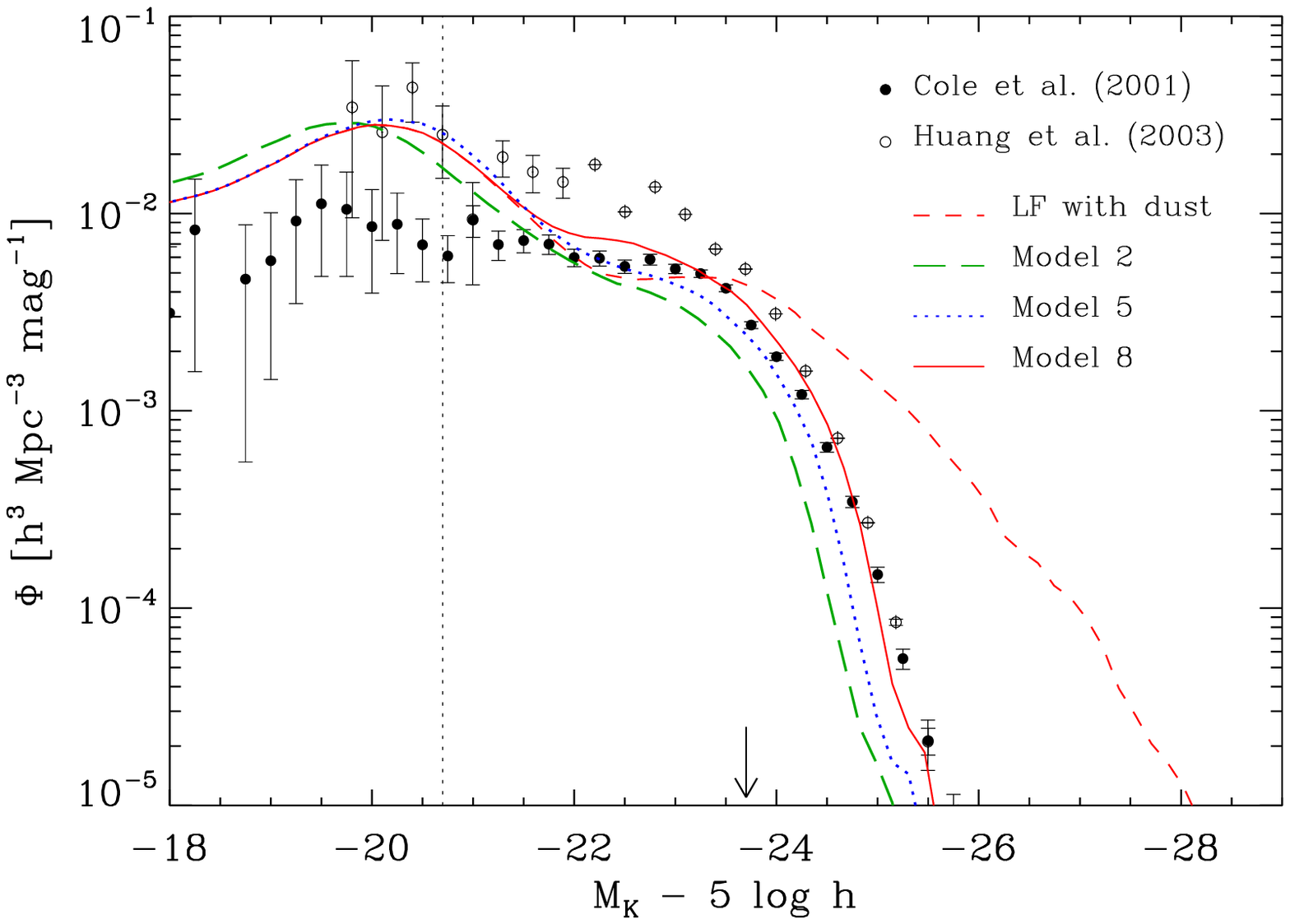}}
{\includegraphics[width=85mm]{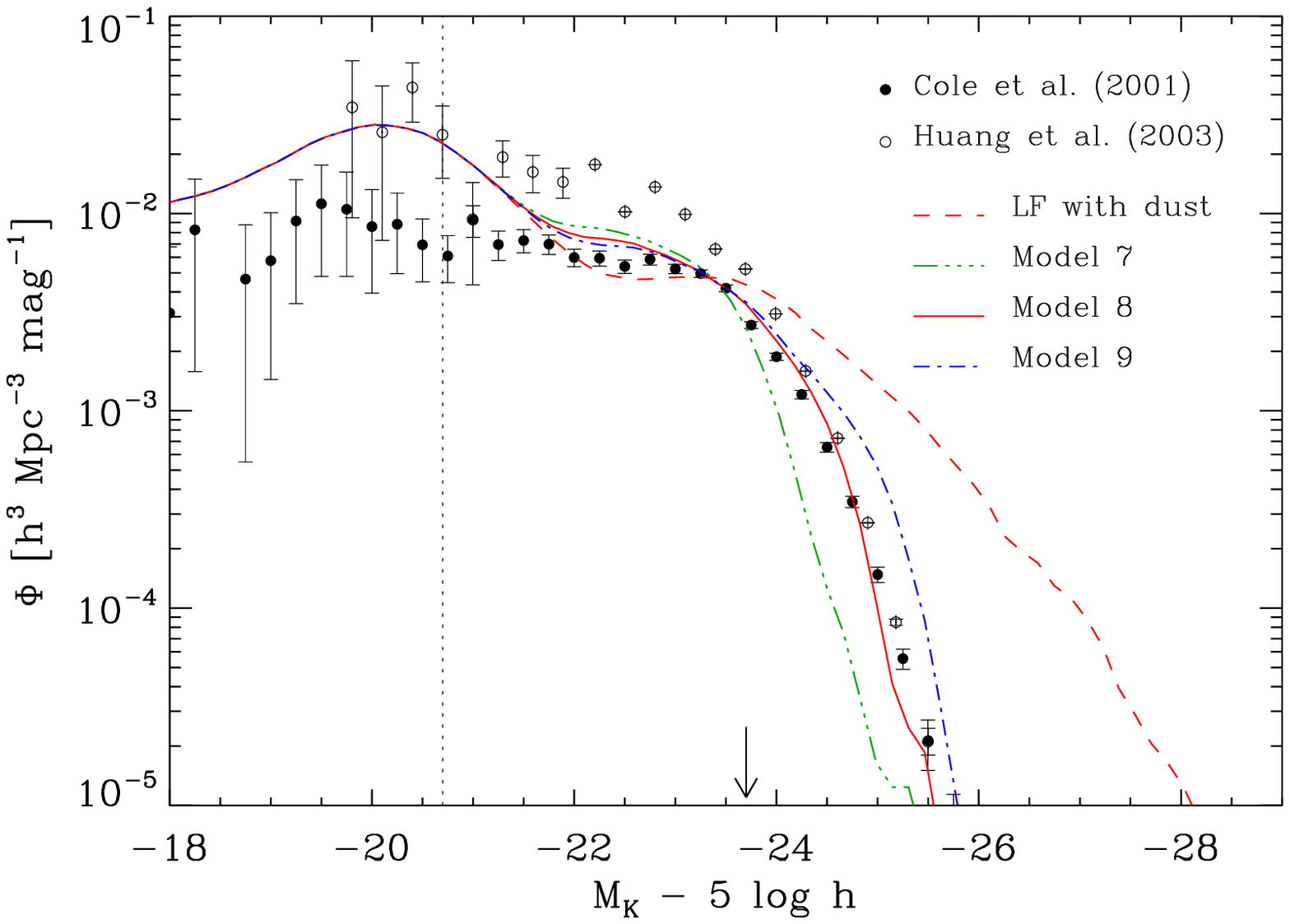}}
\caption{$K$-band field LF at $z=0$ for different SF suppression models (left-hand panel: long-dashed, dotted and 
solid lines which correspond to models 2, 5 and 8 of Table \ref{table1}; right-hand panel: three-dot-dashed, solid and 
dot-dashed lines which correspond to models 7, 8 and 9 of Table \ref{table1}). Solid lines in both panels correspond to 
the {\it best-fitting} model. The simulated LFs are renormalized to match observations at $M_* - 5 \log h$ (indicated by an arrow; see text). 
As a comparison we plot as a dashed line the dust-attenuated LF without SF suppression. The approximate resolution limit of the simulation 
in each band is indicated by the vertical dotted lines.
}
\label{field2}
\end{figure*}

\begin{figure*}
{\includegraphics[width=85mm]{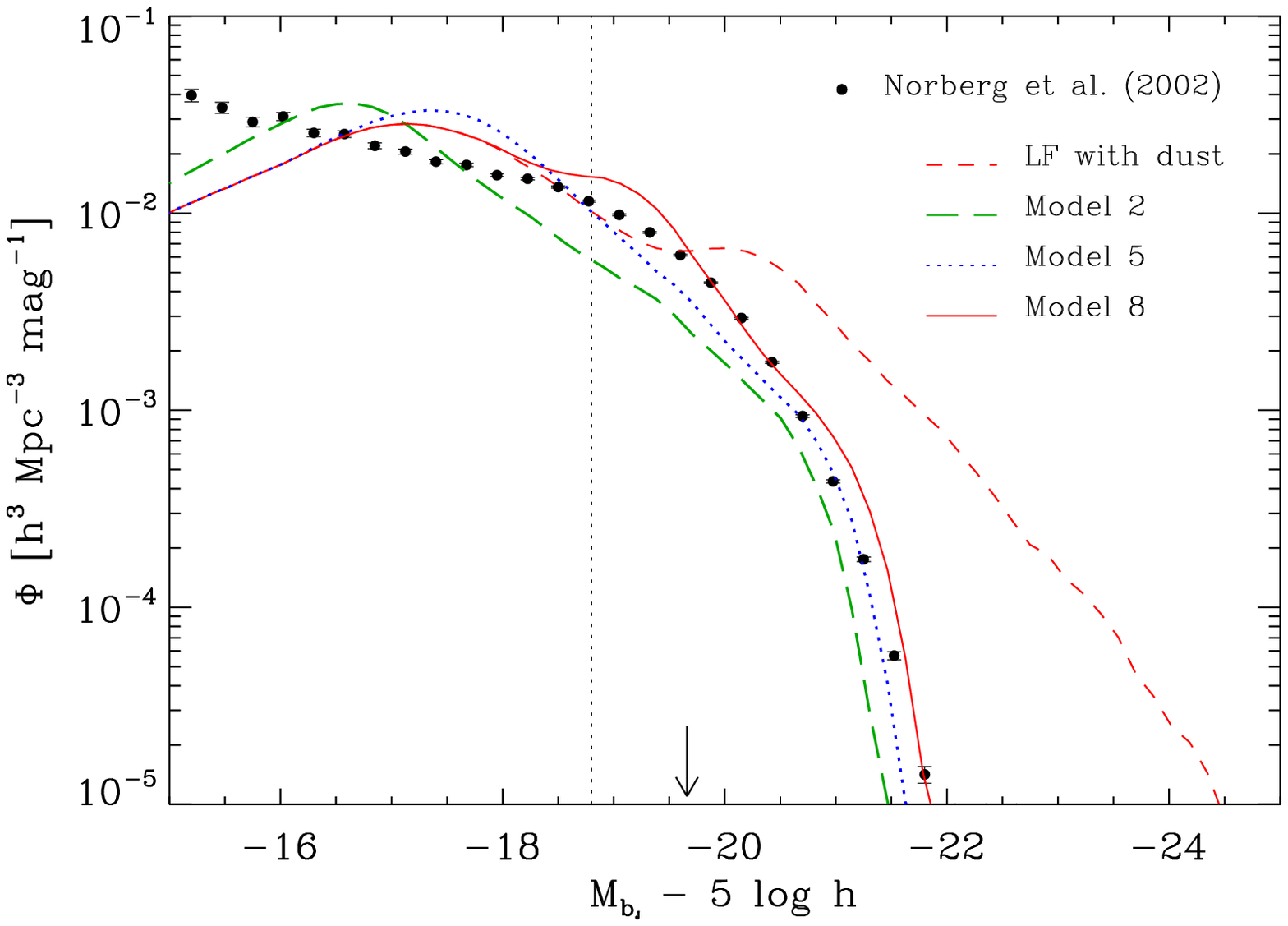}}
{\includegraphics[width=85mm]{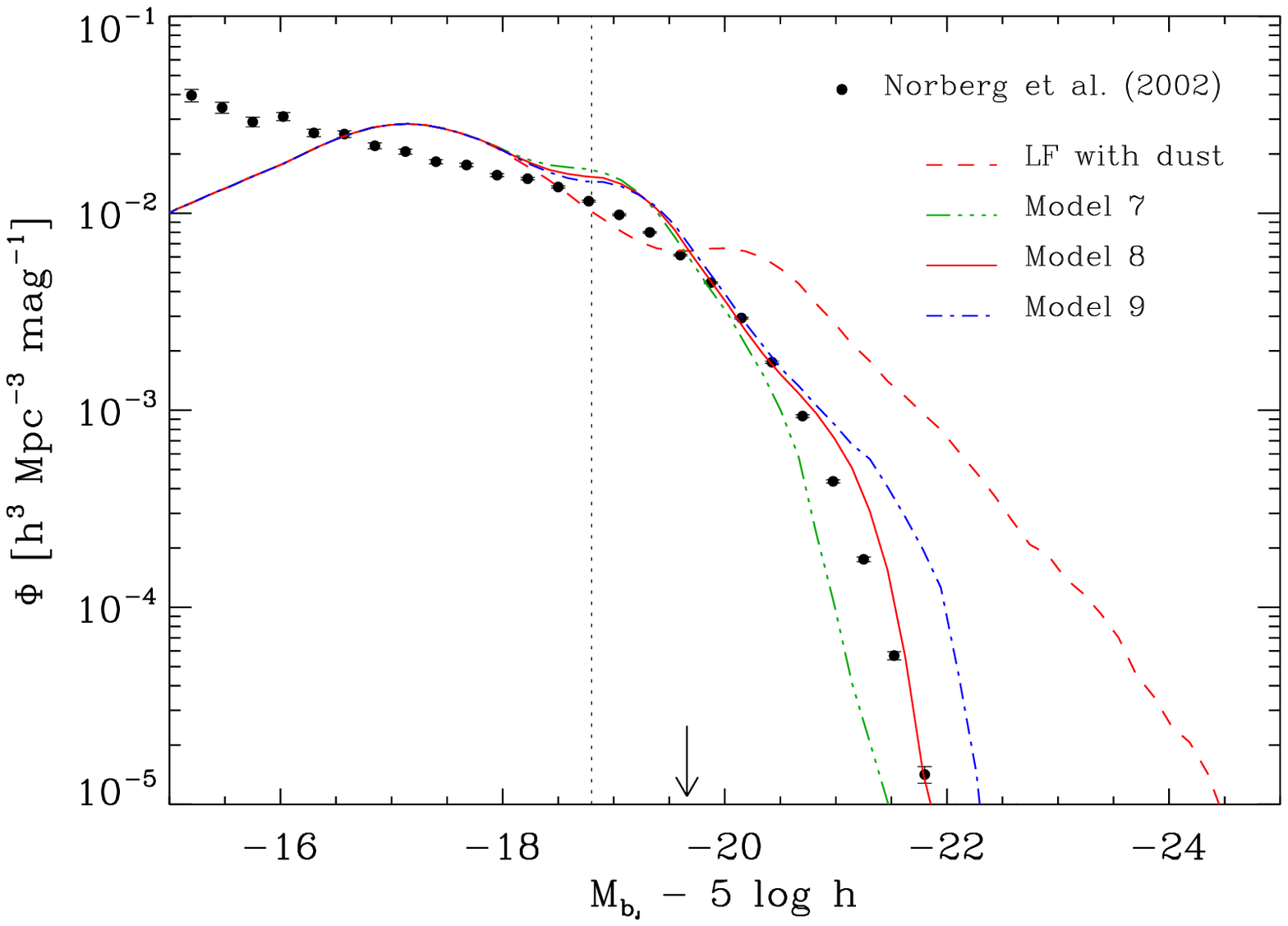}}
\caption{Idem as Fig. \ref{field2} for the $b_{\rm j}$-band field LF.}
\label{field3}
\end{figure*}

\section{Results}
\label{results}

\subsection{Luminosity functions}

In Fig. \ref{field} we plot the $K$- and $b_{\rm j}$-band LFs both for the simulated 
galaxies at $z=0$ and for the observed local ones determined 
by Cole et al. (2001), Huang et al. (2003) and Norberg et al. (2002) respectively. 
In order to match observations at $M_* - 5 \log h$ we renormalized our model LFs in each case. 
In relation to this, we note that the number density of $L_*$ galaxies in our simulation 
gives a value roughly two times smaller than what is observed in each band. 
This fact is also consistent with the number density we obtain for simulated galaxies having a typical 
stellar mass value when comparing with the local stellar mass function (e.g. Li \& White 2009). 
This can be due to several reasons. Firstly, the background cosmology gives a baryon fraction which is smaller 
than what latest observations suggest. Secondly, the underlying numerical resolution adopted here (due to the 
large volume captured) can affect the assembly of small mass systems, thus preventing them of being the building 
blocks of more massive galaxies. These two facts combined could give rise to less galaxies of a 
given stellar mass. However, it is interesting to note that this is in line with the
findings by Saro et al. (2006), where a similar deficit for galaxies within clusters was found, 
i.e. galaxy number above a given magnitude threshold tends to be smaller than observations by a factor of a few.

\begin{figure*}
{\includegraphics[width=88mm]{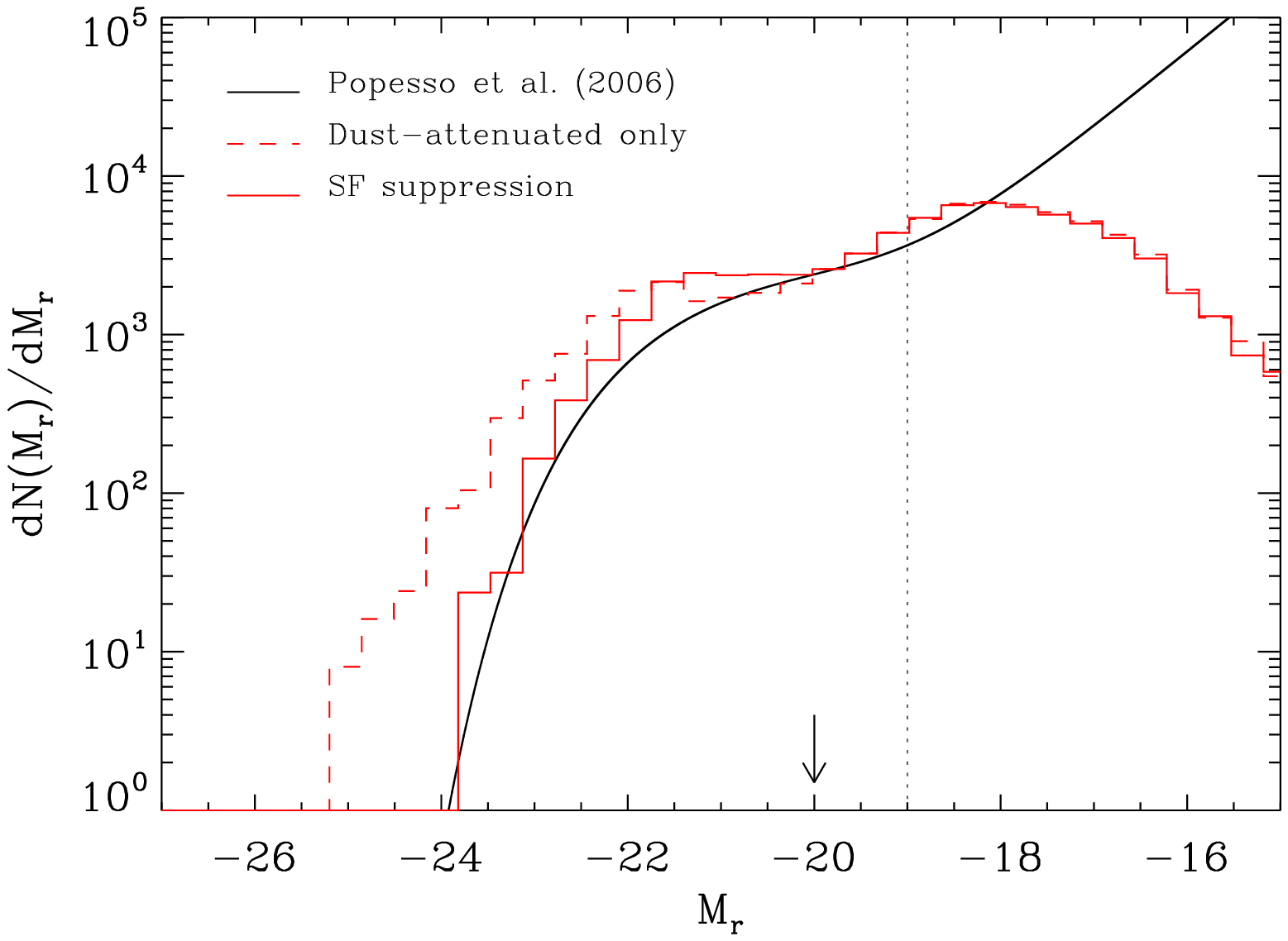}}
{\includegraphics[width=88mm]{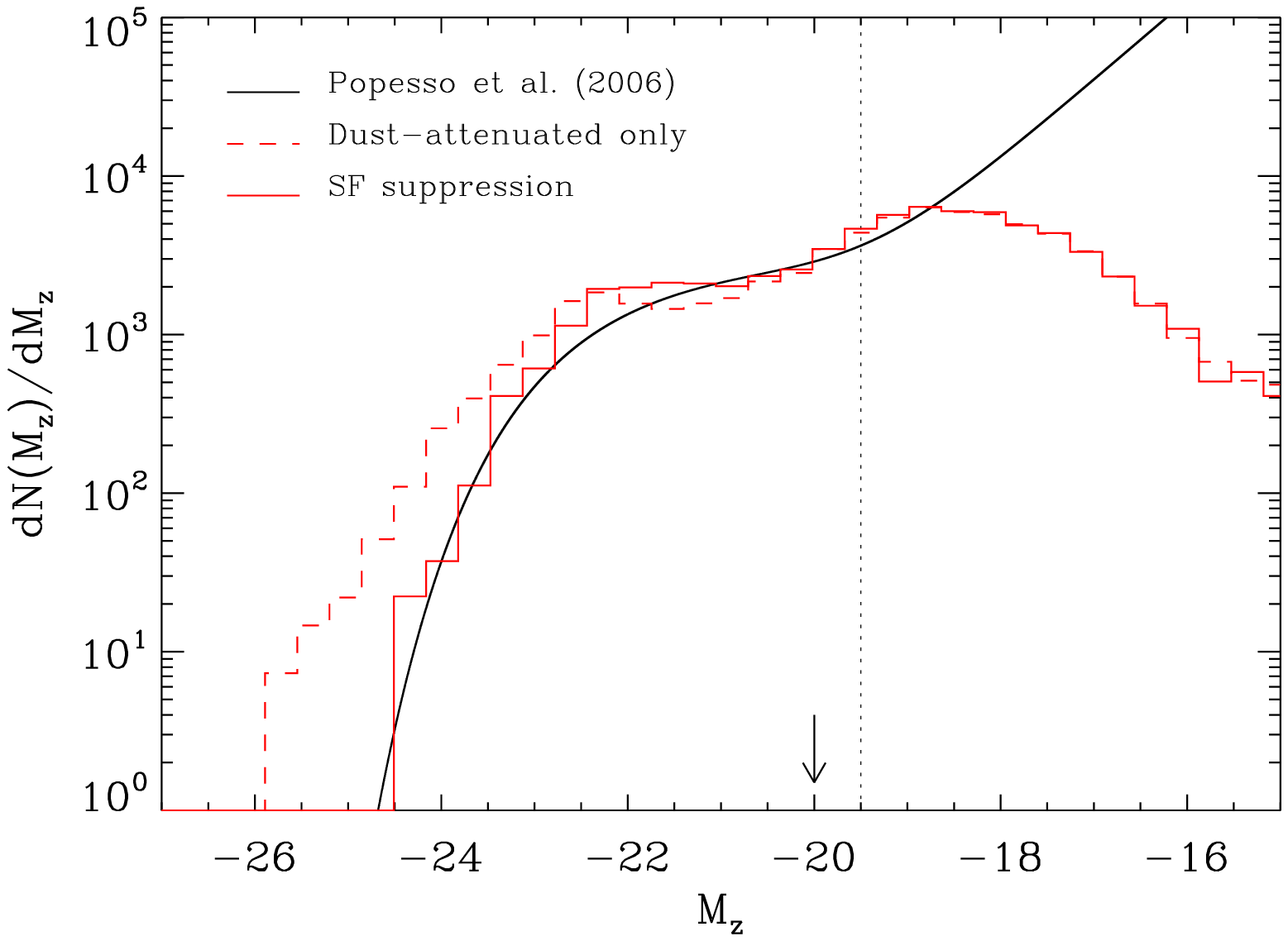}}
\caption{Luminosity distribution for galaxies in groups and clusters for the SDSS $r$- (left-hand panel) 
and $z$-bands (right-hand panel) in the simulation. The best-fitting to data in the RASS-SDSS galaxy cluster 
survey given by Popesso et al. (2006) is also shown as a smooth solid line. The dashed lines correspond to 
simulated cluster galaxies without SF quenching in massive haloes, while the solid ones correspond to 
the {\it best-fitting} SF suppression case. The simulated result is forced to match observations 
at $M_{r,z}=-20$ (indicated by an arrow; see text). The approximate resolution limit of the simulation 
in each band is indicated by the vertical dotted lines.}
\label{cluster}
\end{figure*}

\begin{figure}
{\includegraphics[width=80mm]{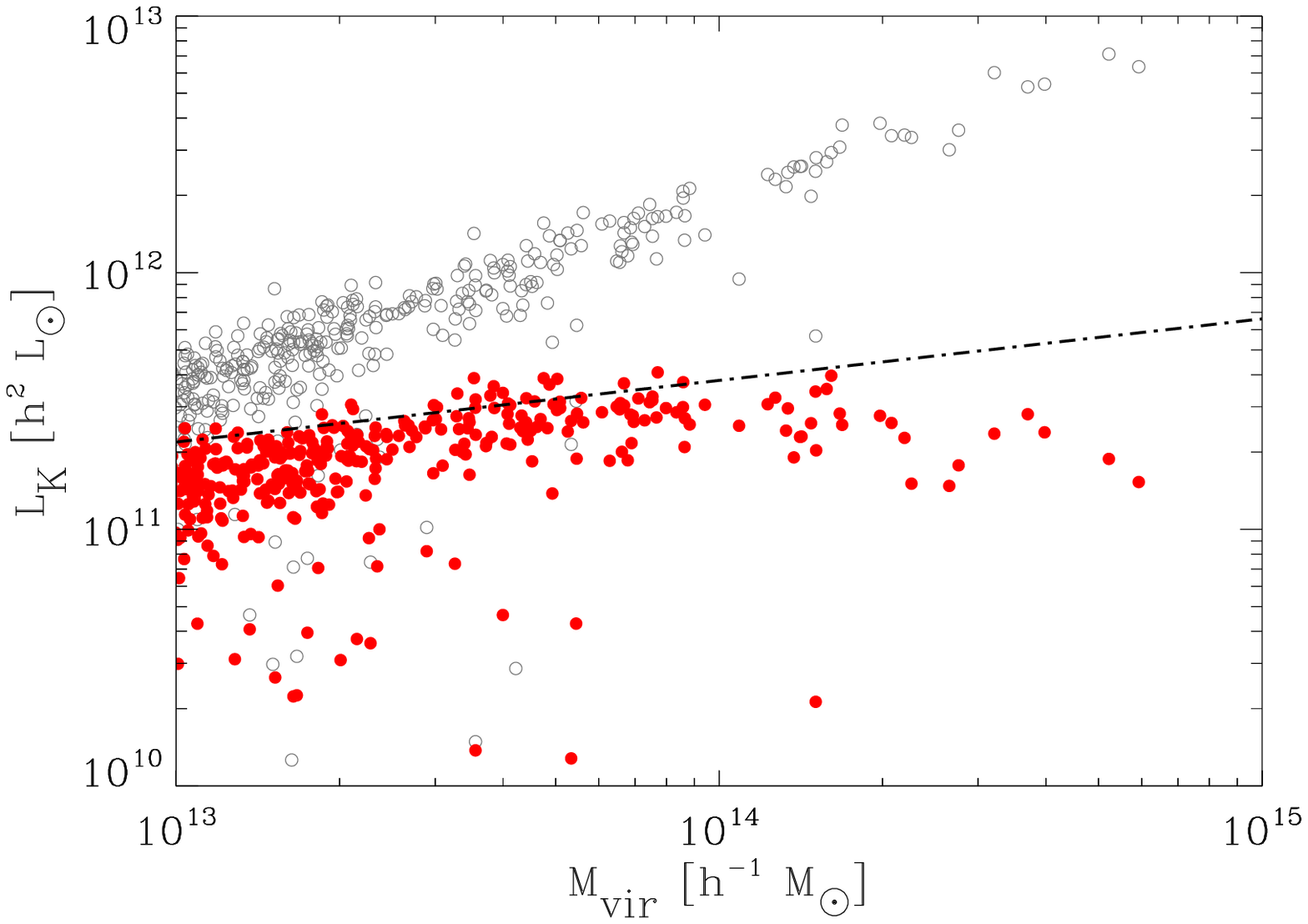}}
\caption{Luminosity-halo mass relation for BCGs in the simulation showing the $K$-band luminosity 
as a function of their virial mass in comparison with the observational trend (dot-dashed line) 
given by Brough et al. (2008). Open circles stand for the SF non-suppression case, while 
filled ones represent our {\it best-fitting} SF quenching scenario.}
\label{LK_MBCG}
\end{figure}

To show the effect of our dust implementation we plot both the attenuated (dashed line) and non-attenuated (solid line) LFs. 
As expected, the dust extinction is more important in the $b_{\rm j}$-band mainly affecting in this case to 
the brightest (and gas-rich) systems. As can be seen in this figure, the simulations are 
not able to reproduce the bright-end of the LFs given by the observations. This is due to gas cooling excess 
onto the dark matter potential wells that leads to an overproduction 
of stars in the most massive objects. As mentioned in the introduction, this behavior was first pointed out 
by SAMs indicating the need of a mechanism capable of supressing SF 
as a function of time and halo mass. Nowadays, there is a growing body of 
evidence, both from theory and observations, that AGN feedback can play an important role in this respect 
(e.g. Bower et al. 2006; Croton et al. 2006; Rafferty et al. 2006; Malbon et al. 2007; Khalatyan et al. 2008; 
Lagos, Cora \& Padilla 2008; Cattaneo et al. 2009), although the subject is far from being completely understood.

In the same spirit of early SAMs (see below), we approximate the SF suppression in massive haloes ignoring a given fraction 
of the formed stars, using the stellar population ages and the maximum circular velocities of the parent 
galaxies present in the $z=0$ selected subhaloes of {\small SUBFIND} as a proxy. The maximum circular velocities serve as 
a measure of subhalo mass and, hence, they can be used to estimate the virial velocity of the systems. The relation between the maximum circular velocity and the virial velocity for haloes in the simulation can be seen in Fig. \ref{vmax_vvir}. 
For each resolved galaxy, we apply a selective cutoff to the stellar populations that were born at different 
epochs to mimic the required SF suppression and test for its effect on the LFs and other observables. 
For the sake of simplicity, we assume a linear relationship between the subhalo maximum circular velocity 
$V_{\rm max}$ and the redshift from which we do not take into account the newly-formed stars $z_{\rm cut}$, 
as follows

\begin{equation}
\label{recipe}
V_{\rm max}={V_0}+z_{\rm cut}{V_1}{\rm ,}
\end{equation}

\noindent where $V_0$ and $V_1$ are free parameters. Although
it is a very simplified parametrization aiming at {\it reconciling} the bright-end's LF problem, 
it should allow to explore the behavior of SF suppression in a simple way. Given a ($V_0,V_1$)
pair, the net result of this recipe is to only affect systems above
the velocity limit given by the zero point of the relation $V_0$,
regulating the {\it strength} of the suppression with its slope
$V_1$. The different models explored for this suppression scheme can
be seen in Table \ref{table1}.

This procedure is similar to what was done in earlier semi-analytic work, where switching off gas cooling in haloes over a certain mass threshold was a common procedure to avoid excessive star formation (e.g. Kauffmann et al. 1999; Hatton et al. 2003; Cora 2006; Cattaneo et al. 2006). Nevertheless, it is fair to note that in our simple scheme, it is not possible to ignore the non-trivial interaction between the non-selected stars in a given halo and the surrounding medium for $z<z_{\rm cut}$, which will mainly affect the column density estimates for the suppressed systems. For instance, the effect of SN feedback due to the ignored stars is able to heat up the surrounding gas lowering the cold gas fraction. In the same way, the presence of these stars in the simulation will also pollute the gas phase with more chemical elements, accordingly affecting the dust properties of the medium. However, within the simple suppression framework adopted here, a more detailed model for dust extinction taking into account these facts is beyond the scope of this paper.

We show the outcome of applying this scheme in Figs. \ref{field2} and \ref{field3} for different parameter choices, where the 
$K$- and $b_{\rm j}$-band LFs can be seen. In the left-hand panels, three different models are shown maintaining $V_1$ fixed at 
200 km s$^{-1}$, whereas $V_0$ adopt the values 100, 150 and 200 km s$^{-1}$ (models 2, 5 and 8 of Table \ref{table1}; 
long-dashed, dotted and solid lines respectively). 
As it can be seen, varying the zero point of the relation produces a horizontal shift in the resulting LF keeping its shape almost unchanged. 
Reducing $V_0$ rises the number of suppressed galaxies thus limiting the number of systems at a given 
luminosity. In the right-hand panels, we also show three different models but fixing $V_0$ at 200 km s$^{-1}$ and 
letting $V_1$ adopt the values 150, 200 and 250 km s$^{-1}$ (models 7, 8 and 9 of Table \ref{table1}; three-dot-dashed, solid and dot-dashed lines respectively). 
In this case, all model LFs coincide at $\sim M_* - 5 \log h$ showing appreciable differences only in the 
bright-end. This reflects the strength of SF suppression recipe in more massive haloes. 

\begin{figure*}
{\includegraphics[width=0.8\textwidth]{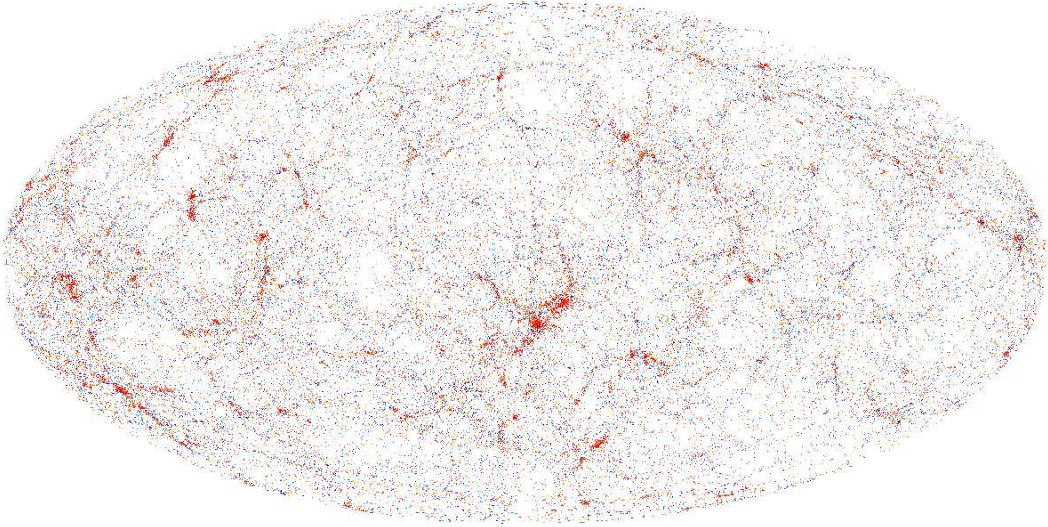}}
\caption{Idem as Fig. \ref{full_sky} but for the galaxy distribution colour-coded from blue to red by 
their $B-V$ colours ($0.3<B-V<1$) in the {\it best-fitting} SF suppression case. Hydrodynamical galaxies 
with redder colours tend to inhabit high density regions in contrast with bluer systems populating 
the field.}
\label{fsgal}
\end{figure*}

For the explored models, the best agreement with observations is obtained for $(V_0,V_1)=(200,200)$ km s$^{-1}$ 
(model 8, see Table \ref{table1}). This maximum circular velocity value roughly corresponds to 
$V_{\rm vir}\sim150$ km s$^{-1}$ as can be seen from Fig. \ref{vmax_vvir}. In this model, systems 
with $V_{\rm vir} \lesssim 150$ km s$^{-1}$ are not affected by the SF quenching and those having 
$V_{\rm vir} \gtrsim 300$ km s$^{-1}$ are suppressed since $z \gtrsim 1$, in agreement with previous SAM 
findings which take into account physically motivated AGN feedback models (e.g. Croton et al. 2006). In what follows, 
we will refer to this case as the {\it best-fitting} model. Even though in this case there is a much better agreement 
with data in the bright-end compared to the non-suppression case, some improvement is still needed in the faint-end 
with the simulation results typically overpredicting the number density of galaxies per magnitude bin in relation to 
observations. As mentioned in the introduction, it is believed that stellar feedback should be more efficient 
in smaller galaxies. Nevertheless, higher resolution simulations are needed 
in order to study this effect.

Given the size of the simulated volume it is possible to study the distribution of luminosities in the groups 
and galaxy clusters formed. 
However, we do not pretend here to make a rigorous comparison with observations since we will address this issue in future 
work using high-resolution runs. Instead, we just select all galaxies belonging to groups with virial mass 
higher than $10^{13}$ $h^{-1}$ M$_{\sun}$ as our galaxy sample. 
In Fig. \ref{cluster} we show the luminosity distribution of this sample for the SDSS $r$- and $z$-bands together with 
the fit to the data given by Popesso et al. (2006) for the RASS-SDSS galaxy cluster survey (smooth solid line). 
To take into account the (arbitrary) adopted normalization in observations our cluster LFs are forced to match the data at $M_{r,z}=-20$. 
As can be seen, the {\it best-fitting} model (solid line) does a better work in describing the observed 
profiles than the dust-attenuated only luminosity distribution (dashed line). This is also true if we study the BCG luminosity 
as a function of halo mass. Fig. \ref{LK_MBCG} shows the $K$-band luminosity-halo mass relation for BCGs residing in haloes with $M_{\rm vir}>10^{13}$ $h^{-1}$ M$_{\sun}$ in both mentioned cases (open and filled circles respectively), compared to the observational trend given by Brough et al. (2008) (dot-dashed line). It can be seen that the suppression scenario displays a better agreement with data, although the suppression seems to be somewhat stronger than needed 
for the most massive systems.

\subsection{Galaxy colours}

In Fig. \ref{fsgal} we show the distribution of galaxies in our simulated local volume projected 
onto the sky. The galaxies are colour coded using their $B-V$ colours in the {\it best-fitting} 
suppression case. From this figure it can be stated qualitatively that redder hydrodynamical galaxies 
tend to inhabit high density environments, in contrast with galaxies populating the field, generally bluer. 
A similar trend is obtained for the plain simulation output, although in that case there is an excess of 
blue systems in cluster environments.  

\begin{figure}
{\includegraphics[width=85mm]{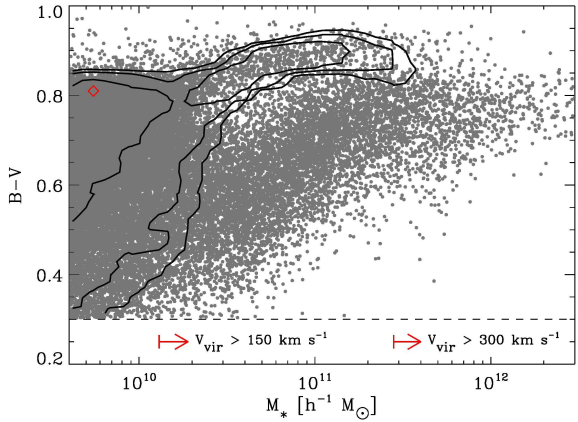}}
\caption{$B-V$ galaxy colours as a function of their stellar mass for the numerically resolved galaxy population at $z=0$. 
Filled circles stand for the case without SF quenching, while contour lines contain 55\%, 85\% and 95\% of the galaxy population 
in the {\it best-fitting} SF suppression scenario. Arrows indicate lower bounds for the virial velocity of systems 
with a given stellar mass: $V_{\rm vir} \gtrsim 150$ and 300 km s$^{-1}$ respectively. 
The diamond indicates the approximate position of the satellite galaxy population in the diagram.}
\label{color_mag}
\end{figure}

The effect of SF suppression in the colours of galaxies is easily seen in Fig. \ref{color_mag}. There we show the $B-V$ 
colour index of numerically resolved galaxies against their stellar masses for the non-suppression case (filled circles), and what 
results for the {\it best-fitting} suppression procedure presented above (contour lines). 
Interestingly, the hydrodynamical galaxy population of the non-suppresed simulation shows a clear colour bi-modality for a given stellar mass, where two main branches can be easily seen. Those systems with colour index satisfying $B-V \gtrsim 0.8$ populate the so-called {\it red sequence}, while the rest populate what we call the {\it blue cloud}. This last branch contains a blue galaxy population with $M_* \gtrsim 2\times10^{11}$ $h^{-1}$ M$_{\odot}$ responsible for the mismatch in the bright-end of the LFs shown in Fig. \ref{field}. 
In particular, most of the systems having $V_{\rm vir} \gtrsim 300$ km s$^{-1}$ (arrow in the right) correspond to overluminous BCGs which have 
acquired around $50\%$ of their stellar mass at late epochs due to the overcooling effect.

This is at variance with the SF suppression scenario (contour lines) in which most of the massive blue galaxies became 
redder as a result of their older stellar populations. 
It can be seen that many of the systems with $V_{\rm vir} \gtrsim 150$ km s$^{-1}$ (arrow in the left) have 
been suppressed in such a way that their colours turn redder than what observations suggest. Henceforth, it is clear from the 
contours that the simple procedure adopted to suppress SF is too efficient for some galaxies, while not efficient enough for the most 
massive ones. This results in a distribution that cannot fully account for the galaxy colour bi-modality in the stellar 
mass range $M_*\sim 2\times10^{10}$ -- $2\times10^{11}$ $h^{-1}$ M$_{\odot}$ 
(e.g. Baldry et al. 2006), reflecting the fact that physically motivated feedback schemes must be 
considered for massive haloes in future work.

As a final remark, it is worth noting that a moderately massive red population having $M_*\sim 5\times10^{9}$ $h^{-1}$ M$_{\odot}$ 
naturally appears in the simulation. The diamond in Fig. \ref{color_mag} indicates the approximate position of this population 
in the diagram. These systems correspond to galaxies that have assembled most of their stellar mass at early epochs, being satellites 
which gas content have been stripped off when accreted by central objects. This mechanism removes gas from galaxies limiting new SF and, 
as a consequence, stellar populations become older and redder. 

\subsection{Clustering properties}

\begin{figure*}
{\includegraphics[width=88mm]{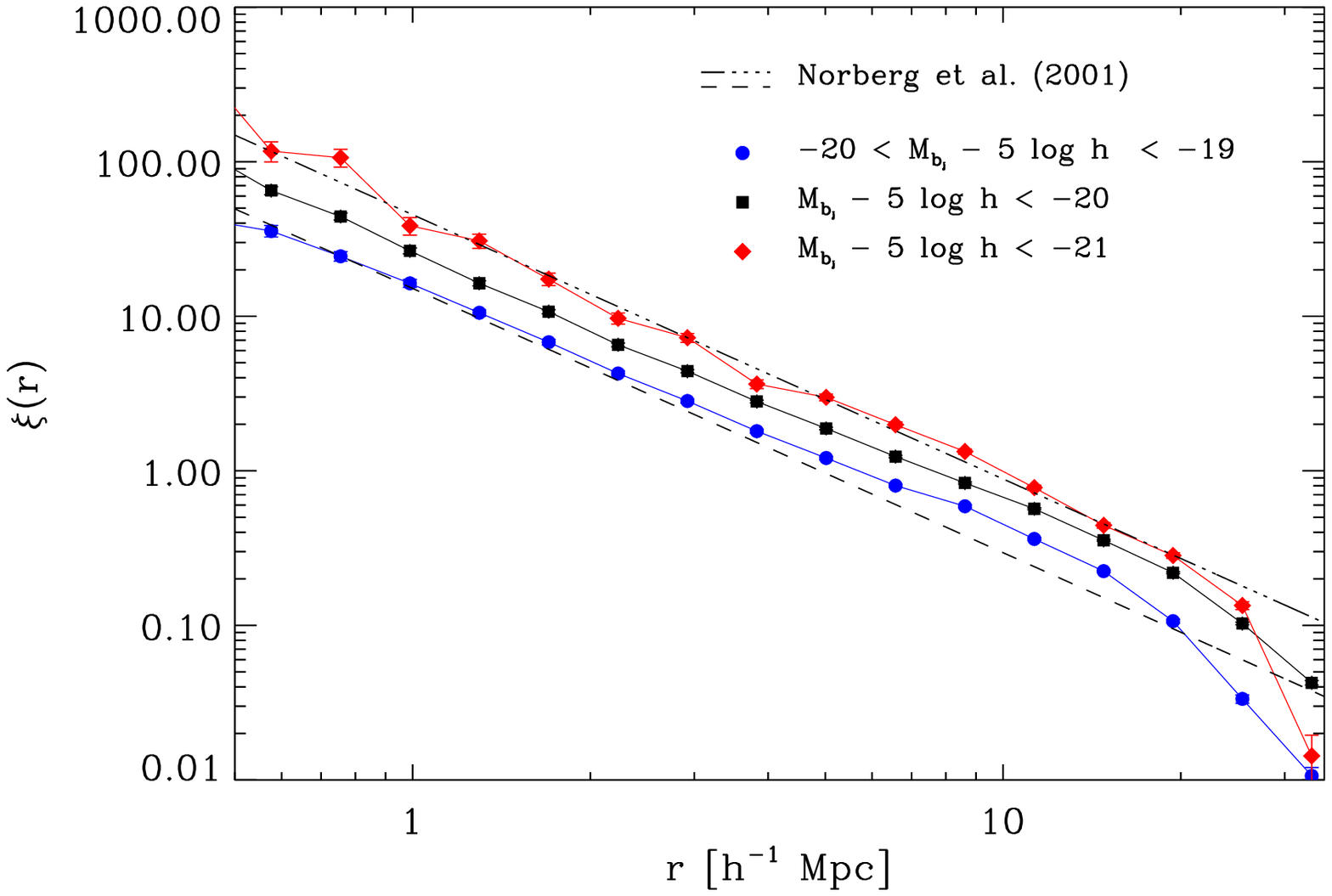}}
{\includegraphics[width=88mm]{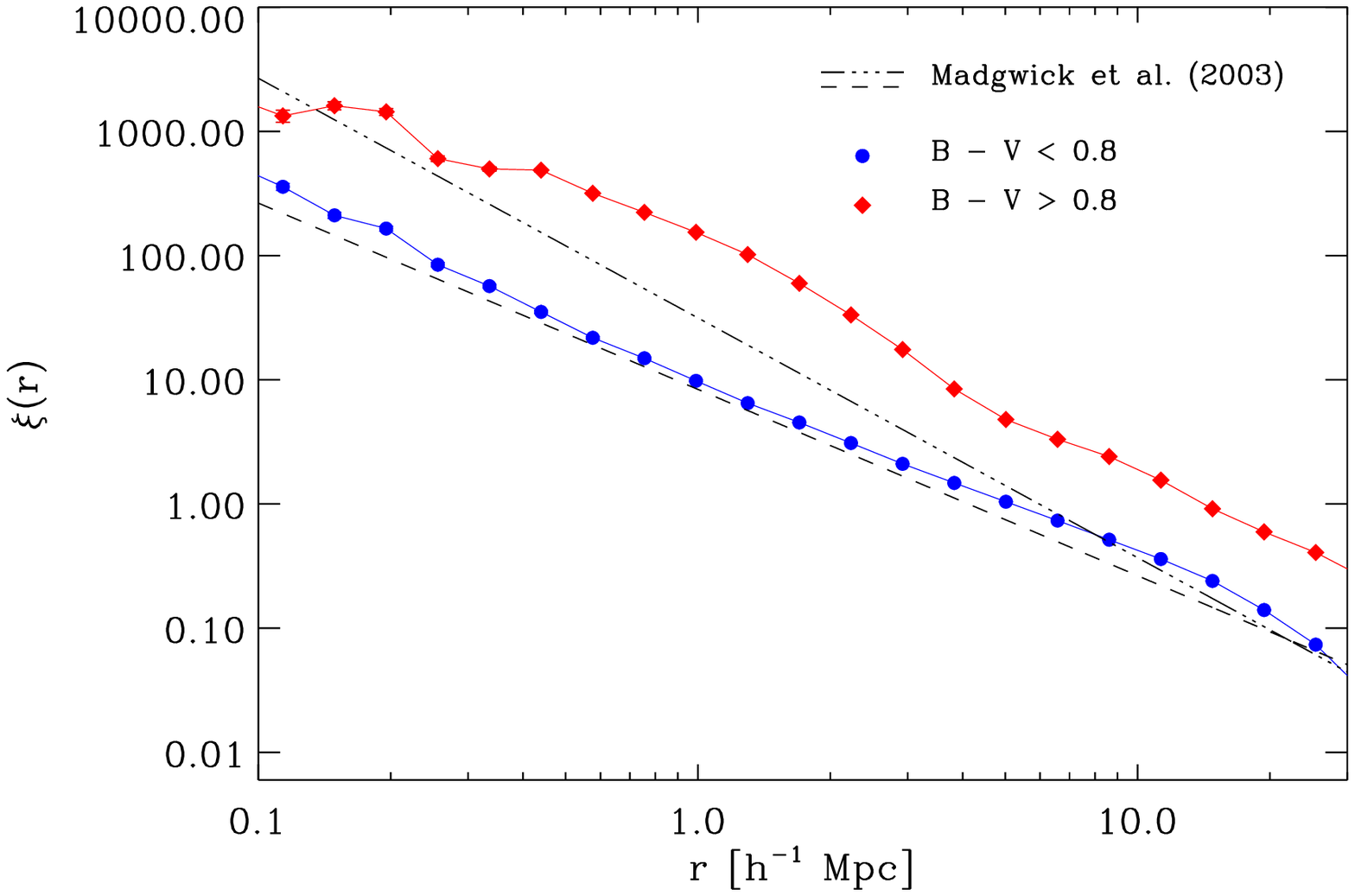}}
\caption{Real space two-point correlation function for the simulated galaxies without SF quenching (filled symbols) 
in different $b_{\rm j}$-band magnitude bins (left-hand panel) and colour populations (right-hand panel). 
The error bars denote Poissonian fluctuations. Observational trends from Norberg et al. (2001) for $\sim L_*$ and 
$\sim 4 L_*$ galaxies, and from Madgwick et al. (2003) for {\it active} and {\it passive} galaxies are shown as 
dashed and three-dot-dashed lines respectively. The observations are valid in the range $\sim0.1$--20$~h^{-1}$ Mpc.}
\label{corr1}
\end{figure*}

In this section, we explore the clustering properties of galaxies in the simulation, both for the 
non-suppresed and {\it best-fitting} SF suppression cases.
The left-hand panel of Fig. \ref{corr1} shows the real-space two-point correlation function for 
galaxies in the plain simulation with different $b_{\rm j}$-band absolute magnitudes (filled symbols). 
Circles are related to galaxies with $-20 < M_{b_{\rm j}} - 5 \log h < -19$, while squares and diamonds show 
the resulting correlations for all galaxies brighter than $-20$ and $-21$, respectively. We also plot as dashed and three-dot-dashed 
lines the observational trends found by Norberg et al. (2001) for $\sim L_*$ and $\sim 4 L_*$ galaxies 
(corresponding to $M_{b_{\rm j}} - 5 \log h \simeq -19.5$ and $-21.5$ respectively). 
In the first place, it can be seen that brighter galaxies cluster strongly than fainter 
ones, as expected. The agreement with observations is fairly good taking into account 
the fact that normalization is a natural outcome of the simulation. 
In particular, for $\sim L_*$ galaxies (circles) the clustering pattern shows a 
power-law behavior similar to that observed, although the agreement is worse for higher scales ($\gtrsim10$ $h^{-1}$ Mpc). 
Likewise, for systems with $M_{b_{\rm j}} - 5 \log h < -21$ (diamonds) the agreement with observations 
is encouraging.

It is also interesting to investigate the clustering of galaxies as a function of their SF 
activity parametrized using galaxy colours. At $z\sim0$, passive systems tend to populate 
high density environments, having redder colours, in contrast to active systems which are mainly located 
in the field (e.g. Balogh et al. 2004; Baldry et al. 2004, 2006). 
We define as {\it passive} systems all galaxies having $B-V>0.8$, while {\it active} systems satisfy $B-V<0.8$. 
The right-hand panel of Fig. \ref{corr1} shows the real-space two-point correlation function for passive red galaxies 
(diamonds) and bluer ones (circles) in comparison with the observed 
correlations (valid for scales between $\sim0.1$--20$~h^{-1}~\rm{Mpc}$) given by Madgwick et al. (2003). 
It is easy to see that both populations are clearly separated, with redder systems clustering more strongly, and 
displaying a similar behavior than observations. For passive systems, there is a substantial excess of power 
and a clear departure from a power-law. The power excess in this case is due to the fact that the 
overcooling present in the simulation makes the number density of red systems low, thus rising the value of the 
correlation length.

\begin{figure*}
{\includegraphics[width=88mm]{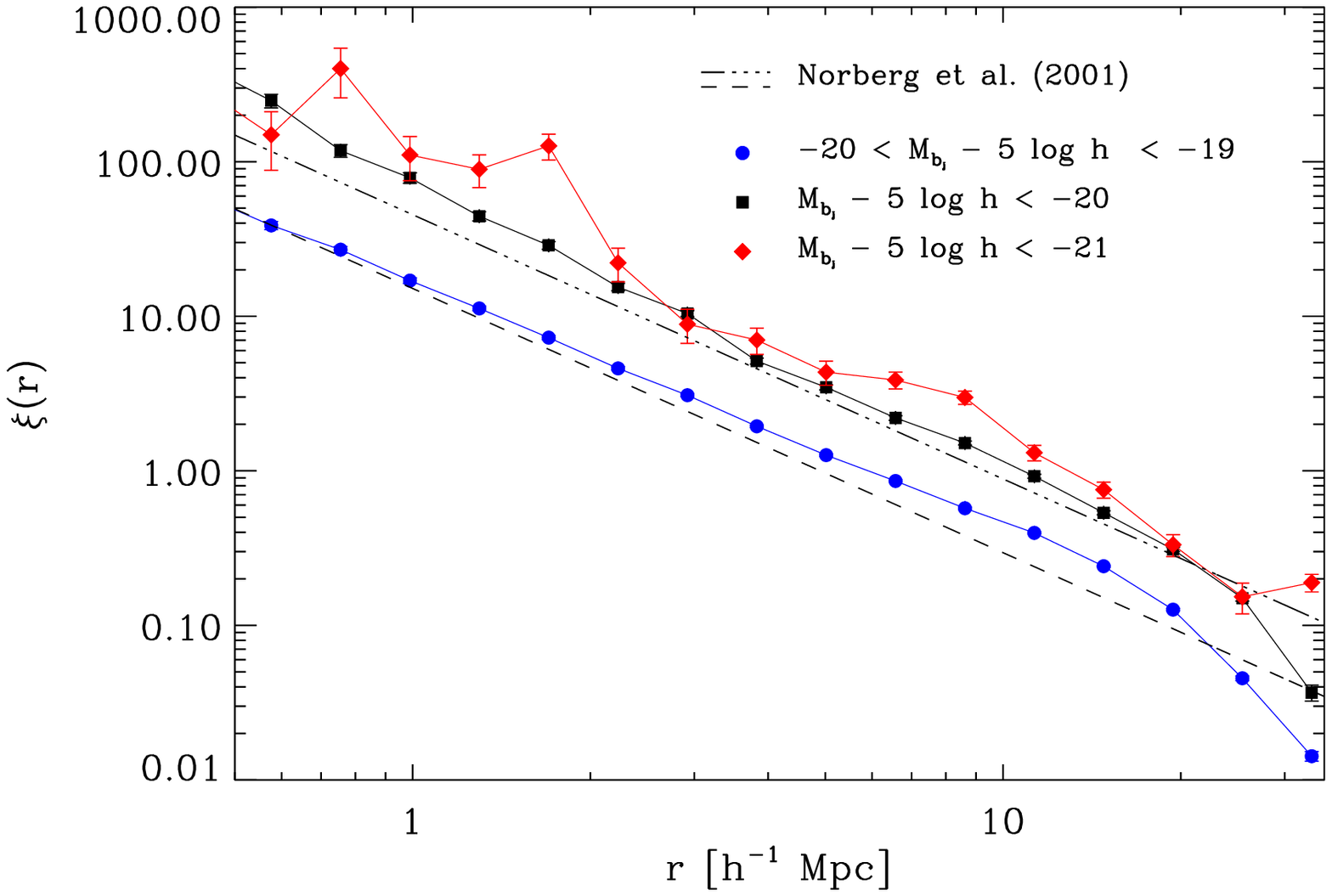}}
{\includegraphics[width=88mm]{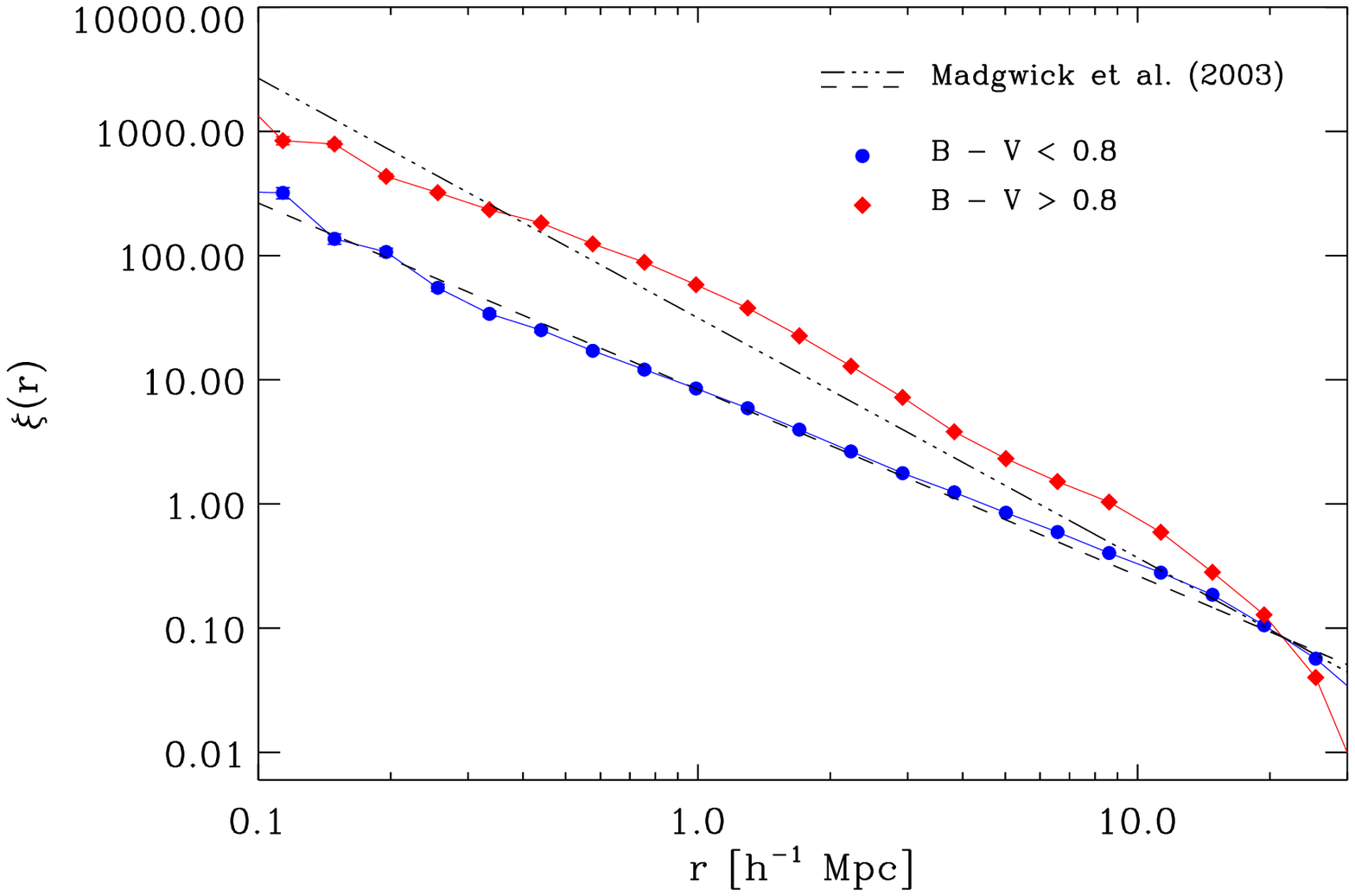}}
\caption{Idem as Fig. \ref{corr1} for galaxies in the {\it best-fitting} post-processing scheme to supress SF.}
\label{corr2}
\end{figure*}

On the other hand, Fig. \ref{corr2} shows the outcomes when considering the {\it best-fitting} suppression scheme. 
From its left-hand panel, it can be seen that the correlation function for $\sim L_*$ galaxies (circles) 
exhibits essentially the same behavior as before (compare with Fig. \ref{corr1}), while displaying appreciable 
differences only for brighter galaxies (squares and diamonds). For objects brighter than $M_{b_{\rm j}} - 5 \log h = -21$ 
the result is affected by low number statistics leading to a noisier correlation function that is somewhat off from the observational 
expectation. This is partially overcome when including more galaxies in the $M_{b_{\rm j}} - 5 \log h > -20$ case.
Active galaxies present an excellent agreement with data for all studied scales (right-hand panel of Fig. \ref{corr2}), and the 
discrepancy with observations found for passive red systems found in Fig. \ref{corr1} is alleviated. This is somewhat expected 
since, as can be seen from the contour lines of Fig. \ref{color_mag}, after the SF quenching, many of the galaxies have migrated to the 
{\it red sequence} rising the number density of passive systems. Despite of this improvement, it is important to remember that the 
colour-stellar mass relation is not well reproduced in this case due to the simplistic approach adopted 
to {\it redden} galaxies more massive than $V_{\rm vir}\gtrsim 150$ km s$^{-1}$.

\section{Summary and Conclusions}
\label{concl}

In this paper, we have focused on the luminosity, colour and clustering properties of galaxies 
that arise naturally in a hydrodynamical cosmological simulation of structure formation. The 
spherical local volume of $\sim130^3$ $h^{-3}$ Mpc$^3$ was simulated using the {\it zooming} technique 
with more than $\sim 50$ million gas and dark matter particles. In particular, 
our simulation code include different physical mechanisms such as supernova feedback, galactic 
winds, an UV pervading background since $z\sim6$ and chemical enrichment of the gas and stellar populations as time 
elapses. However, we have not explicitly included a physical mechanism capable of quenching SF in massive 
haloes (like AGNs). In this way, by $z=0$ a total of $\sim20000$ objects having more 
than 32 star particles are formed in the volume. The luminosities of the hydrodynamical galaxies 
were computed using the differential contribution provided by every star particle present inside the optical radius of the systems. 
As a way of estimating the dust-attenuated luminosity in the bands of interest, we implemented a simple 
model for dust extinction that makes use of the cold gas distribution in each galaxy as a tracer of 
its dust content. In order to mimic the required SF suppression in massive 
(i.e. $V_{\rm vir} \gtrsim 150$ km s$^{-1}$) haloes we also applied a simple recipe to the present day output of the simulation 
to reconcile the bright-end of the simulated LFs with observations and to further investigate its inclusion 
on other properties of the simulated galaxies. In the following we summarize the main 
conclusions of the present work:

\begin{itemize}
\item As shown earlier by semi-analytic work, it is essential to suppress SF in massive haloes to 
describe the bright-end of the field LF at $z\sim0$. For instance, haloes with masses corresponding to 
$V_{\rm vir}\sim300$ km s$^{-1}$ are required to quench their SF since $z\sim1$ in the {\it best-fitting} 
SF suppression case. In this way, it is possible to avoid the formation of very bright blue BCGs in the 
simulation. However, after a small renormalization to $L_*$, the faint-end of the simulated field LFs tends 
to overpredict the number density of galaxies per magnitude bin in the $b_{\rm j}$-band for galaxies with 
luminosities lower than $\sim L_*$. To better study this effect higher resolution simulations are needed.\\

\item The group/cluster bright-end luminosity distributions are better described when SF suppression is considered. 
This is due to the fact that massive cluster satellites used to be BCGs in the past, and also require some level 
of SF suppression. However, as a consequence of our relatively high-mass galaxy resolution limit we cannot 
confirm the upturn of faint cluster satellites in the cluster LF function as previously obtained by Saro et al. (2006, 2009) 
using high-resolution resimulations of galaxy clusters. On the other hand, the $K$-band luminosity-halo mass relation that 
naturally arises in the simulation also needs further SF quenching in massive haloes to better describe observations 
given by Brough et al. (2008).\\

\item The stellar masses of the hydrodynamical galaxies in the plain simulation display a clear $B-V$ colour 
bi-modal behavior as suggested by observations (e.g. Baldry et al. 2006), although the {\it active} blue galaxies present 
very massive systems, due to overcooling, contrary to what data shows. 
When taking into account the SF suppression scheme, the $B-V$ galaxy colours turn redder, displaying a 
colour-stellar mass distribution that is not able to fully account for observations. 
In this case, a bi-modal distribution is observed for stellar masses lower than $\sim2\times10^{10}$ $h^{-1}$ M$_{\odot}$, while for 
higher masses most of the galaxies are red. This is a manifestation of the rough nature of the simple parametrization adopted 
to quench SF in massive haloes.\\

\item The clustering properties of different hydrodynamical galaxy populations that naturally arise in the simulation 
are in qualitative agreement with observational results. The scale-dependent correlation function of our $\sim L_*$ 
objects in the $b_{\rm j}$-band displays a power-law behavior up to $\sim10~h^{-1}$ Mpc that is in good agreement with 
data given by Norberg et al. (2001). 
As expected, higher luminosity systems show a stronger clustering pattern, where galaxies having luminosities around 
$\sim 4L_*$ display an excellent agreement with observations. On the other hand, it is also possible to disentangle the 
galaxy correlation properties when the sample is splitted by galaxy colours. The correlation function of {\it active} 
blue ($B-V<0.8$) systems is in rough agreement with the results given by Madgwick et al. (2003). However, 
{\it passive} red systems ($B-V>0.8$) present an excess of power in comparison with observations, showing a clear departure 
from the power-law behavior. As shown when using the SF suppression scheme, this discrepancy is alleviated when more galaxies are 
able to populate the {\it red sequence} of the colour-magnitude diagram. This is due to the fact that, typically, the added 
systems tend to reside in less clustered haloes, thus lowering the correlation length value.\\

In summary, we conclude that analyzing the global properties of the galaxy population within hydrodynamical,
cosmological simulations, start to be a promising tool to study galaxy evolution. Current simulations already 
fairly represent the underlying hydrodynamical effects (at least in a global sense) and, in general, describe the star formation 
process well enough to qualitatively reproduce observed environmental trends. However, similar than for the widely 
used SAMs, overcooling in  massive halos has to be quenched by additional feedback effects. It has to be seen in 
future simulations, which directly include such additional processes, if such a quenching happens mildly enough not 
to destroy some of general trends already captured in the current generation of hydrodynamical simulations.

\end{itemize}

\section*{Acknowledgments}
The authors would like to thank the referee, Adrian Jenkins, for several comments that helped to improve 
this paper. S. E. N.~acknowledges G. De Lucia, E. Puchwein, A. G. S\'anchez, F. Stasyszyn and C. Scannapieco for useful 
discussions and DAAD (Deutscher Akademischer Austausch Dienst) for support. K.D.~acknowledges support by the DFG Priority 
Programme 1177.


\begin{thebibliography}{}

\bibitem{baldry04}
Baldry I. K., Glazebrook K., Brinkmann J., Ivezi\'c, Z., Lupton R. H., Nichol R. C., Szalay A. S., 
2004, ApJ, 600, 681
\bibitem{baldry06}
Baldry I. K., Balogh M. L., Bower R. G., Glazebrook K., Nichol R. C., Bamford S. P., Budavari T., 
2006, MNRAS, 373, 469
\bibitem{balogh04}
Balogh M. L., Baldry I. K., Nichol R., Miller C., Bower R., Glazebrook K., 2004, ApJ, 615, 101
\bibitem{blanton01}
Blanton M. R. et al., 2001, AJ, 121, 2358
\bibitem{blanton03}
Blanton M. R. et al., 2003, ApJ, 592, 819
\bibitem{baugh95}
Baugh C. M., Gazta\~naga E., Efstathiou G., 1995, MNRAS, 274, 1049
\bibitem{bower06}
Bower R. G., Benson A. J., Malbon R., Helly J. C., Frenk C. S., Baugh C. M., Cole S., Lacey C. G., 2006, MNRAS, 370, 645
\bibitem{Brough08}
Brough S., Couch W. J., Collins C. A., Jarrett T., Burke D. J., Mann R. G., 2008, MNRAS, 385, 103
\bibitem{bc03}
Bruzual G., Charlot S., 2003, MNRAS, 344, 1000
\bibitem{cattaneo06}
Cattaneo A., Dekel A., Devriendt J., Guiderdoni B., Blaizot J., 2006, MNRAS, 370, 1651
\bibitem{cattaneo09}
Cattaneo A. et al., 2009, Nat, 460, 213
\bibitem{charlot00}
Charlot S., Fall S. M., 2000, ApJ, 539, 718
\bibitem{cole01}
Cole S. et al., 2001, MNRAS, 326, 255
\bibitem{crain09}
Crain R. A. et al., 2009, MNRAS, 399, 1773
\bibitem{croton06}
Croton D. J. et al., 2006, MNRAS, 365, 11
\bibitem{colless99}
Colless M., 1999, R. Soc. London Philos. Trans. A, 357, 1750, 105
\bibitem{colless01}
Colless M. et al., 2001, MNRAS, 328, 1039
\bibitem{cora06}
Cora S., 2006, MNRAS, 368, 1540
\bibitem{do07}
Dav\'e R., Oppenheimer B. D., 2007, MNRAS, 374, 427
\bibitem{delucia04}
De Lucia G., Kauffmann G., White S. D. M., 2004, MNRAS, 349, 1101
\bibitem{delucia07}
De Lucia G., Blaizot J., 2007, MNRAS, 375, 2
\bibitem{dolag05}
Dolag, K., Hansen F. K., Roncarelli M., Moscardini L., 2005, MNRAS, 363, 29
\bibitem{dolag09}
Dolag K., Borgani S., Murante G., Springel V., 2009, MNRAS, 399, 497
\bibitem{fisher94}
Fisher K. B., Davis M., Strauss M. A., Yahil A., Huchra J., 1994, MNRAS, 266, 50
\bibitem{fisher95}
Fisher K. B., Huchra J. P., Strauss M. A., Davis M., Yahil A., Schlegel D., 1995, ApJS, 100, 69
\bibitem{fontanot09}
Fontanot F., De Lucia G., Monaco P., Somerville R., Santini P., 2009, MNRAS, 397, 1776
\bibitem{guiderdoni87}
Guiderdoni B., Rocca-Volmerange B., 1987, A\&A, 186, 1
\bibitem{guo_white}
Guo Q., White S. D. M., 2009, MNRAS, 396, 39
\bibitem{haardt96}
Haardt F., Madau P., 1996, ApJ, 461, 20
\bibitem{hatton03}
Hatton S., Devriendt J.E.G., Ninin S., Bouchet F. R., Guiderdoni B., Vibert D., 2003, MNRAS, 343, 75
\bibitem{hawkins03}
Hawkins E. et al., 2003, MNRAS, 346, 78
\bibitem{hoffman91}
Hoffman Y., Ribak E., 1991, ApJ, 380, 5
\bibitem{huang03}
Huang J. S., Glazebrook K., Cowie L. L., Tinney C., 2003, ApJ, 584, 203
\bibitem{kauffmann99}
Kauffmann G., Colberg J. M., Diaferio A., White S. D. M., 1999, MNRAS, 303, 188
\bibitem{khalatyan08}
Khalatyan A., Cattaneo A., Schramm M., Gottl\"ober S., Steinmetz M., Wisotzki L., 2008, MNRAS, 387, 13
\bibitem{kolatt96}
Kolatt T., Dekel A., Ganon G., Willick J. A., 1996, ApJ, 458, 419
\bibitem{lagos06}
Lagos C. Del P., Cora S. A., Padilla N. D., 2008, MNRAS, 388, 587
\bibitem{Li09}
Li C., White S. D. M., 2009, MNRAS, 398, 2177L
\bibitem{Lin96}
Lin H., Kirshner R. P., Shectman S. A., Landy S. D., Oemler A., Tucker D. L., Schechter P. L., 1996, ApJ, 464, 60
\bibitem{madgwick03}
Madgwick D. S. et al., 2003, MNRAS, 344, 847
\bibitem{maeder89}
Maeder A., Meynet G., 1989, A\&A, 210, 155
\bibitem{malbon07}
Malbon R. K., Baugh C. M., Frenk C. S., Lacey C. G., 2007, MNRAS, 382, 1394
\bibitem{matteucci01}
Matteucci F., Recchi S., 2001, ApJ, 558, 351
\bibitem{mathis02}
Mathis H., Lemson G., Springel V., Kauffmann G., White S. D. M., Eldar A., Dekel A., 2002, 333, 739
\bibitem{norberg01}
Norberg P. et al., 2001, MNRAS, 328, 64
\bibitem{norberg02}
Norberg P. et al., 2002, MNRAS, 336, 907
\bibitem{nuza07}
Nuza S. E., Tissera P. B., Pellizza L. J., Lambas D. G., Scannapieco C., de Rossi M. E., 2007, MNRAS, 375, 665
\bibitem{ocvirk08}
Ocvirk P., Pichon C., Teyssier R., 2008, MNRAS, 390, 1326
\bibitem{od06}
Oppenheimer B. D., Dav\'e R., 2006, MNRAS, 373, 1265
\bibitem{od08}
Oppenheimer B. D., Dav\'e R., 2008, MNRAS, 387, 577
\bibitem{pearce01}
Pearce F. R., Jenkins A., Frenk C. S., White S. D. M., Thomas P. A., Couchman H. M. P., Peacock J. A., 
Efstathiou G., 2001, MNRAS, 326, 649
\bibitem{popesso04}
Popesso P., B\"ohringer H., Brinkmann J., Voges W., York D. G., 2004, A\&A, 423, 449
\bibitem{popesso06}
Popesso P., Biviano A., B\"ohringer H., Romaniello M., 2006, A\&A, 445, 29
\bibitem{recchi01}
Recchi S., Matteucci F., D'Ercole A., 2001, MNRAS, 322, 800
\bibitem{rafferty06}
Rafferty D. A., McNamara B. R., Nulsen P. E. J., Wise M. W., 2006, ApJ, 652, 216
\bibitem{salpeter55}
Salpeter E. E., 1955, ApJ, 121, 161
\bibitem{saro06}
Saro A., Borgani S., Tornatore L., Dolag K., Murante G., Biviano A., Calura F., Charlot S., 2006, MNRAS, 373, 397
\bibitem{saro08}
Saro A., De Lucia G., Dolag K., Borgani S., 2008, MNRAS, 391, 565
\bibitem{saro09}
Saro A., Borgani S., Tornatore, L., De Lucia G., Dolag K., Murante G., 2009, MNRAS, 392, 795
\bibitem{spergel03}
Spergel D.N. et al., 2003, ApJS, 148, 175
\bibitem{springel01}
Springel V., White S. D. M., Tormen G., Kauffmann G., 2001, MNRAS, 328, 726
\bibitem{springel02}
Springel V., Hernquist L., 2002, MNRAS, 333, 649
\bibitem{springel03}
Springel V., Hernquist L., 2003, MNRAS, 339, 289
\bibitem{springel05a}
Springel V., 2005, MNRAS, 364, 1105
\bibitem{springel05b}
Springel V. et al., 2005, Nat, 435, 629
\bibitem{sd93}
Sutherland R. S., Dopita M. A., 1993, ApJS, 88, 253
\bibitem{tornatore04}
Tornatore L., Borgani S., Matteucci F., Recchi S., Tozzi P., 2004, MNRAS, 349, 19
\bibitem{tornatore07}
Tornatore L., Borgani S., Dolag K., Matteucci F., 2007, MNRAS, 382, 1050
\bibitem{white01}
White M., Hernquist L., Springel V., 2001, preprint (astro-ph/0107023)
\bibitem{york00}
York D. G. et al., 2000, AJ, 120, 1579
\bibitem{yoshikawa01}
Yoshikawa K., Taruya A., Jing Y.P., Suto Y., 2001, ApJ, 558, 520
\bibitem{zehavi}
Zehavi I. et al., 2002, ApJ, 571, 172


\end{thebibliography}
\end{document}